%Paper: hep-ph/9207223
%From: PHI091@ibm.southampton.ac.uk
%Date: Wed, 08 Jul 92 19:15:07 BST

%
\magnification=1200
\hsize=31pc \vsize=55 truepc
\hfuzz=2pt \vfuzz=4pt
\pretolerance=5000 \tolerance=5000
\parskip=0pt plus 1pt \parindent=16pt
%
% fonts for title
\font\fourteenrm=cmr10 scaled \magstep2
\font\fourteeni=cmmi10 scaled \magstep2
\font\fourteenbf=cmbx10 scaled \magstep2
\font\fourteenit=cmti10 scaled \magstep2
\font\fourteensy=cmsy10 scaled \magstep2
% font for small caps within title and authors names
\font\large=cmbx10 scaled \magstep1
%
% font for matrices, please replace with cmbx10
% if you do not have this font available

% for matrices, use \bss{x} for bold sans serif within maths

%
% font for vectors (bold italic), please replace with cmbx10
% if you do not have this font available

% for vectors, use \bi{r} for bold italic r within maths

%
% fonts for small type (if used)
%
\font\eightrm=cmr8
\font\eighti=cmmi8
\font\eightbf=cmbx8
\font\eightit=cmti8

\font\eightsy=cmsy8
\font\sixrm=cmr6
\font\sixi=cmmi6
\font\sixsy=cmsy6

\def\tenpoint{\def\rm{\fam0\tenrm}%
  \textfont0=\tenrm \scriptfont0=\sevenrm
                      \scriptscriptfont0=\fiverm
  \textfont1=\teni  \scriptfont1=\seveni
                      \scriptscriptfont1=\fivei
  \textfont2=\tensy \scriptfont2=\sevensy
                      \scriptscriptfont2=\fivesy
  \textfont3=\tenex   \scriptfont3=\tenex
                      \scriptscriptfont3=\tenex
  \textfont\itfam=\tenit  \def\it{\fam\itfam\tenit}%
  \textfont\slfam=\tensl  \def\sl{\fam\slfam\tensl}%
  \textfont\bffam=\tenbf  \scriptfont\bffam=\sevenbf
                            \scriptscriptfont\bffam=\fivebf
                            \def\bf{\fam\bffam\tenbf}%
  \normalbaselineskip=16 truept
  \setbox\strutbox=\hbox{\vrule height14pt depth6pt width0pt}%
  \let\sc=\eightrm \normalbaselines\rm}
\def\eightpoint{\def\rm{\fam0\eightrm}%
  \textfont0=\eightrm \scriptfont0=\sixrm
                      \scriptscriptfont0=\fiverm
  \textfont1=\eighti  \scriptfont1=\sixi
                      \scriptscriptfont1=\fivei
  \textfont2=\eightsy \scriptfont2=\sixsy
                      \scriptscriptfont2=\fivesy
  \textfont3=\tenex   \scriptfont3=\tenex
                      \scriptscriptfont3=\tenex
  \textfont\itfam=\eightit  \def\it{\fam\itfam\eightit}%
  \textfont\bffam=\eightbf  \def\bf{\fam\bffam\eightbf}%
  \normalbaselineskip=13 truept
  \setbox\strutbox=\hbox{\vrule height11pt depth5pt width0pt}}
\def\fourteenpoint{\def\rm{\fam0\fourteenrm}%
  \textfont0=\fourteenrm \scriptfont0=\tenrm
                      \scriptscriptfont0=\eightrm
  \textfont1=\fourteeni  \scriptfont1=\teni
                      \scriptscriptfont1=\eighti
  \textfont2=\fourteensy \scriptfont2=\tensy
                      \scriptscriptfont2=\eightsy
  \textfont3=\tenex   \scriptfont3=\tenex
                      \scriptscriptfont3=\tenex
  \textfont\itfam=\fourteenit  \def\it{\fam\itfam\fourteenit}%
  \textfont\bffam=\fourteenbf  \scriptfont\bffam=\tenbf
                             \scriptscriptfont\bffam=\eightbf
                             \def\bf{\fam\bffam\fourteenbf}%
  \normalbaselineskip=19 truept
  \setbox\strutbox=\hbox{\vrule height17pt depth7pt width0pt}%
  \let\sc=\tenrm \normalbaselines\rm}

\def\monthyear{\ifcase\month\or
  January\or February\or March\or April\or May\or June\or
  July\or August\or September\or October\or November\or
December\fi
  \space \number\year}
%
% counter definitions
%
\newcount\secno      %section number
\newcount\subno      %number of subsection
\newcount\subsubno   %number of subsubsection
\newcount\appno      %appendix number
\newcount\tableno    %table number
\newcount\figureno   %figure number
\normalbaselineskip=16 truept
\baselineskip=16 truept
\footline={\ifnum\pageno=0 \hfil \else
\hss\tenrm\number\pageno\hss \fi}
%
% Title of article
\def\title#1
   {\pageno=0\vglue1truein
   {\baselineskip=19 truept
    \pretolerance=10000
    \raggedright
    \noindent \fourteenpoint\bf #1\par}
    \vskip1truein minus36pt}
%
% Names of all the authors in form initials then surname,
% no points after initials, surname in upper and lower case
\def\author#1
  {{\pretolerance=10000
    \raggedright
    \noindent {\large #1}\par}}
%
% Address(es) of author(s)
% if differing addresses use one \address for each
\def\address#1
   {\bigskip
    \noindent \rm #1\par}
%
% short title (not more than fifty characters)
\def\shorttitle#1{}
%
% Physics Abstracts classification numbers
\def\pacs#1{\def\pacsZZZ{{\noindent \rm PACS number(s): #1\par}}}
%
% Journal article submitted to
\def\jnl#1{\def\jnlZZZ{{\noindent \rm Submitted to: {\sl
#1}\par}}}
%
% Month and year
\def\date{\def\dateZZZ{{\noindent Date: \monthyear\par}}}
%
% Start of abstract
\def\beginabstract
   {\vfill
    \noindent {\bf Abstract. }\rm}
%
% Keyword abstract - only required for J. Phys. G
\def\keyword#1
   {\bigskip
    \noindent {\bf Keyword abstract: }\rm#1}
%
% End of abstract
\def\endabstract
   {\par
    \vfill\pacsZZZ\jnlZZZ\dateZZZ\eject}
%
% Contents page only required for Reports on Progress in Physics
% Heading for contents page

%
% entry in list of contents (section headings)
\def\entry#1#2#3
   {\noindent
    \hangindent=20pt
    \hangafter=1
    \hbox to20pt{#1 \hss}#2\hfill #3\par}
%
% subentry in list of contents (subsection heading)
% subsubsection headings do not appear in the contents list
\def\subentry#1#2#3
   {\noindent
    \hangindent=40pt
    \hangafter=1
    \hskip20pt\hbox to20pt{#1 \hss}#2\hfill #3\par}
%
% Section heading (#1 is title of section, no number required)
\def\section#1
   {\vskip0pt plus.1\vsize\penalty-250
    \vskip0pt plus-.1\vsize\vskip24pt plus12pt minus6pt
    \subno=0 \subsubno=0
    \global\advance\secno by 1
    \noindent {\bf \the\secno. #1\par}
    \bigskip
    \noindent}
%
% Subsection (#1 is title of subsection, no number required)
\def\subsection#1
   {\vskip-\lastskip
    \vskip24pt plus12pt minus6pt
    \bigbreak
    \global\advance\subno by 1
    \subsubno=0
    \noindent {\sl \the\secno.\the\subno. #1\par}
    \nobreak
    \medskip
    \noindent}
%
% Heading for list of figure captions

%
% Figure caption, #1 is caption, no number required
\def\figcaption#1
   {\global\advance\figureno by 1
    \noindent {\bf Figure \the\figureno.} \rm#1\par
    \bigskip}
%
% Heading for list of references
\def\references
     {\vfill\eject
     {\noindent \bf References\par}
      \parindent=0pt
      \bigskip}
%
% reference to a journal article in Harvard (alphabetical) system
\def\refjl#1#2#3#4
   {\hangindent=16pt
    \hangafter=1
    \rm #1
   {\frenchspacing\sl #2
    \bf #3}
    #4\par}
%
% reference to a book or report in Harvard (alphabetical) system
\def\refbk#1#2#3
   {\hangindent=16pt
    \hangafter=1
    \rm #1
   {\frenchspacing\sl #2}
    #3\par}
%
% reference to a journal article in numerical system
\def\numrefjl#1#2#3#4#5
   {\parindent=40pt
    \hang
    \noindent
    \rm {\hbox to 30truept{\hss #1\quad}}#2
   {\frenchspacing\sl #3\/
    \bf #4}
    #5\par\parindent=16pt}
%
% reference to a book or report in numerical system
\def\numrefbk#1#2#3#4
   {\parindent=40pt
    \hang
    \noindent
    \rm {\hbox to 30truept{\hss #1\quad}}#2
   {\frenchspacing\sl #3\/}
    #4\par\parindent=16pt}
%
% dash for use with repeated authors in reference lists

%
% Renaming the dot under macro

%
% \d now used for differential d in mathematics
\def\d{{\rm d}}
%
% \e gives roman e for exponential e in mathematics

%
% \i gives roman i for square root of minus one in maths mode
% and dotless i in text mode
\def\i{\ifmmode{\rm i}\else\char"10\fi}
%
% small (text size) fraction within displayed mathematics

%
% et al
\def\etal{{\sl et al\/}\ }
%
% special macros for display equations
%
\catcode`\@=11
%
% \ind defines the 5pc indentation for displayed maths
\def\ind{\hbox to 5pc{}}
%
% \eq(#1) will give the equation number (#1) on the right
% instead of \eqno
\def\eq(#1){\hfill\llap{(#1)}}
\catcode`\@=12
%
% Macro for special accented characters
% vectors with hats

% vectors with overbar
\def\bbar#1{\hbox{\bf\={\bdi #1}}}
% roman characters with right pointing arrow

%
% abbreviations for IOPP journals
%

        %1968-87
   %1988 and onwards
     %1968--1988
        %1989 and onwards

           %1975--1988
\def\jpg{J. Phys. G: Nucl. Part. Phys.}     %1989 and onwards

%
% Other commonly quoted journals
%

\def\NC{Nuovo Cim.}

\def\NP{Nucl. Phys.}
\def\PL{Phys. Lett.}
\def\PR{Phys. Rev.}

%
% Article begins here
%
\def\up{\uparrow}\def\dn{\downarrow}
\def\Lqcd{\Lambda_{\rm QCD}}
\def\as{\alpha_{\rm s}}
\def\su#1{SU(#1)}
\def\bra#1{\left\langle #1 \right|}
\def\ket#1{\left| #1 \right\rangle}
\def\vv{\vec{v}}\def\pv{\vec{p}}
\def\pvp{\vec{p}^{\,\prime}}
\def\SL{\vec{S}_\ell}
\def\slt{s_\ell}
\def\half{{1\over2}}
\def\Qbar{\smash{\overline{Q}}\vphantom{Q}}
\def\Bbar{\smash{\overline{B}}\vphantom{B}}
\def\bbar{\overline{b}}\def\cbar{\overline{c}}
\def\g#1{\gamma^{#1}}
\def\hqgamma{{\rlap{I}\kern1.3pt\Gamma}} % Gamma for HQET current
\def\enu{e \overline{\nu}}
\def\labelto#1{\quad{\buildrel \displaystyle #1
\over\longrightarrow}\quad}
% following two \def's for Feynman slash
\def\cancel#1#2{\ooalign{$\hfil#1\mkern1mu/\hfil$\crcr$#1#2$}}
\def\slash#1{\mathpalette\cancel{#1}}
\def\vproj{{1+\slash v\over2}}
\def\propline{\hrule width60pt depth0.5pt height0.5pt}
\def\prop{\vcenter{\propline}}
\def\heavyprop{\vcenter{\propline\kern0.8pt\propline}}
\def\ept{\eightpoint}
\def\it{\sl}
%
%
% \jlref\label{text} or \bkref\label{text}
% Use for journals and books/preprints/conf-proceedings
% respectively.  Generates a number, assigns it to \label,
% generates an entry, puts [#] in text.  Generate reference
% without [#] in text with \njlref and \nbkref.  Cite
% reference by [\label] in the text. To list the refs,
% \listrefs
%
\global\newcount\refno \global\refno=1
\newwrite\rfile
\def\jlref{[\the\refno]\njlref}
\def\njlref#1{\xdef#1{\the\refno}\checknstep
\immediate\write\rfile{\noexpand\numrefjl{[#1]}}\findarg}
\def\bkref{[\the\refno]\nbkref}
\def\nbkref#1{\xdef#1{\the\refno}\checknstep
\immediate\write\rfile{\noexpand\numrefbk{[#1]}}\findarg}
\def\checknstep{\ifnum\refno=1\immediate\openout
\rfile=\jobname.ref\fi\global\advance\refno by1
\chardef\wfile=\rfile}
%
% horrible hack to sidestep tex \write limitation
\def\findarg#1#{\begingroup\obeylines%
\newlinechar=`\^^M\passarg}
{\obeylines\gdef\passarg#1{\writeline\relax #1^^M\hbox{}^^M}%
\gdef\writeline#1^^M{\expandafter\toks0\expandafter%
{\striprelax #1}%
\edef\next{\the\toks0}\ifx\next\null\let\next=\endgroup%
\else\ifx\next\empty%
\else\immediate\write\wfile{\the\toks0}\fi%
\let\next=\writeline\fi\next\relax}}
{\catcode`\%=12\xdef\pctsign{%}}\def\striprelax#1{}
\def\listrefs{\immediate\closeout\rfile%
\references\input\jobname.ref}
%
% following macros to draw circle with blob in centre
% \point from TeXbook p 389
\newdimen\unit
\def\point#1 #2 #3{\rlap{\kern#1\unit
  \raise#2\unit\hbox to 0pt{\hss$#3$\hss}}}
\def\cpt#1 #2 {\point #1 #2 \cdot}
\newbox\circle
% save a typeset circle
\setbox\circle=\hbox{\unit=50pt
\cpt 1.0000 0.5000 \cpt 0.9990 0.5314 \cpt 0.9961 0.5627 \cpt
0.9911 0.5937 \cpt 0.9843 0.6243 \cpt 0.9755 0.6545 \cpt 0.9649
0.6841 \cpt 0.9524 0.7129 \cpt 0.9382 0.7409 \cpt 0.9222 0.7679
\cpt 0.9045 0.7939 \cpt 0.8853 0.8187 \cpt 0.8645 0.8423 \cpt
0.8423 0.8645 \cpt 0.8187 0.8853 \cpt 0.7939 0.9045 \cpt 0.7679
0.9222 \cpt 0.7409 0.9382 \cpt 0.7129 0.9524 \cpt 0.6841 0.9649
\cpt 0.6545 0.9755 \cpt 0.6243 0.9843 \cpt 0.5937 0.9911 \cpt
0.5627 0.9961 \cpt 0.5314 0.9990 \cpt 0.5000 1.0000 \cpt 0.4686
0.9990 \cpt 0.4373 0.9961 \cpt 0.4063 0.9911 \cpt 0.3757 0.9843
\cpt 0.3455 0.9755 \cpt 0.3159 0.9649 \cpt 0.2871 0.9524 \cpt
0.2591 0.9382 \cpt 0.2321 0.9222 \cpt 0.2061 0.9045 \cpt 0.1813
0.8853 \cpt 0.1577 0.8645 \cpt 0.1355 0.8423 \cpt 0.1148 0.8187
\cpt 0.0955 0.7939 \cpt 0.0778 0.7679 \cpt 0.0619 0.7409 \cpt
0.0476 0.7129 \cpt 0.0351 0.6841 \cpt 0.0245 0.6545 \cpt 0.0157
0.6244 \cpt 0.0089 0.5937 \cpt 0.0039 0.5627 \cpt 0.0010 0.5314
\cpt 0.0000 0.5000 \cpt 0.0010 0.4686 \cpt 0.0039 0.4374 \cpt
0.0089 0.4063 \cpt 0.0157 0.3757 \cpt 0.0245 0.3455 \cpt 0.0351
0.3160 \cpt 0.0476 0.2871 \cpt 0.0618 0.2591 \cpt 0.0778 0.2321
\cpt 0.0955 0.2061 \cpt 0.1147 0.1813 \cpt 0.1355 0.1577 \cpt
0.1577 0.1355 \cpt 0.1813 0.1148 \cpt 0.2061 0.0955 \cpt 0.2321
0.0778 \cpt 0.2591 0.0619 \cpt 0.2871 0.0476 \cpt 0.3159 0.0351
\cpt 0.3455 0.0245 \cpt 0.3756 0.0157 \cpt 0.4063 0.0089 \cpt
0.4373 0.0039 \cpt 0.4686 0.0010 \cpt 0.5000 0.0000 \cpt 0.5314
0.0010 \cpt 0.5626 0.0039 \cpt 0.5937 0.0089 \cpt 0.6243 0.0157
\cpt 0.6545 0.0245 \cpt 0.6840 0.0351 \cpt 0.7129 0.0476 \cpt
0.7408 0.0618 \cpt 0.7679 0.0778 \cpt 0.7939 0.0955 \cpt 0.8187
0.1147 \cpt 0.8422 0.1355 \cpt 0.8645 0.1577 \cpt 0.8852 0.1813
\cpt 0.9045 0.2061 \cpt 0.9221 0.2321 \cpt 0.9381 0.2591 \cpt
0.9524 0.2871 \cpt 0.9649 0.3159 \cpt 0.9755 0.3455 \cpt 0.9843
0.3756 \cpt 0.9911 0.4063 \cpt 0.9961 0.4373 \cpt 0.9990 0.4686
}
% use the circle in \muck
\def\muck#1{\vcenter{%
 \hbox{\copy\circle\unit=50pt
 \point 0.5 0.5 \bullet
 \point 0.5 0.3 {#1}}}\hbox to 50pt{\hfil}}
\newbox\stretch
% stretched muck
\setbox\stretch=\hbox{\unit=50pt
\cpt 0.8536 0.8536 \cpt 0.8307 0.8751 \cpt 0.8065 0.8951 \cpt
0.7810 0.9135 \cpt 0.7545 0.9304 \cpt 0.7270 0.9455 \cpt 0.6986
0.9589 \cpt 0.6694 0.9704 \cpt 0.6395 0.9801 \cpt 0.6091 0.9880
\cpt 0.5782 0.9938 \cpt 0.5471 0.9978 \cpt 0.5157 0.9998 \cpt
0.4843 0.9998 \cpt 0.4530 0.9978 \cpt 0.4218 0.9938 \cpt 0.3909
0.9880 \cpt 0.3605 0.9801 \cpt 0.3306 0.9704 \cpt 0.3014 0.9589
\cpt 0.2730 0.9455 \cpt 0.2455 0.9304 \cpt 0.2190 0.9135 \cpt
0.1936 0.8951 \cpt 0.1694 0.8751 \cpt 0.1465 0.8536 \cpt 0.1250
0.8307 \cpt 0.1049 0.8065 \cpt 0.0865 0.7811 \cpt 0.0696 0.7545
\cpt 0.0545 0.7270 \cpt 0.0411 0.6986 \cpt 0.0296 0.6694 \cpt
0.0199 0.6395 \cpt 0.0120 0.6091 \cpt 0.0062 0.5782 \cpt 0.0022
0.5471 \cpt 0.0002 0.5157 \cpt 0.0002 0.4843 \cpt 0.0022 0.4530
\cpt 0.0062 0.4218 \cpt 0.0120 0.3909 \cpt 0.0198 0.3605 \cpt
0.0296 0.3306 \cpt 0.0411 0.3014 \cpt 0.0545 0.2730 \cpt 0.0696
0.2455 \cpt 0.0864 0.2190 \cpt 0.1049 0.1936 \cpt 0.1249 0.1694
\cpt 0.1464 0.1465 \cpt 0.1693 0.1250 \cpt 0.1935 0.1049 \cpt
0.2189 0.0865 \cpt 0.2455 0.0696 \cpt 0.2730 0.0545 \cpt 0.3014
0.0411 \cpt 0.3306 0.0296 \cpt 0.3605 0.0199 \cpt 0.3909 0.0120
\cpt 0.4218 0.0062 \cpt 0.4529 0.0022 \cpt 0.4843 0.0002 \cpt
0.5157 0.0002 \cpt 0.5470 0.0022 \cpt 0.5782 0.0062 \cpt 0.6090
0.0120 \cpt 0.6395 0.0198 \cpt 0.6693 0.0295 \cpt 0.6985 0.0411
\cpt 0.7270 0.0545 \cpt 0.7545 0.0696 \cpt 0.7810 0.0864 \cpt
0.8064 0.1049 \cpt 0.8306 0.1249 \cpt 0.8536 0.1464 \cpt 0.8765
0.1249 \cpt 0.9007 0.1049 \cpt 0.9261 0.0865 \cpt 0.9526 0.0696
\cpt 0.9801 0.0545 \cpt 1.0085 0.0411 \cpt 1.0377 0.0296 \cpt
1.0676 0.0199 \cpt 1.0980 0.0120 \cpt 1.1289 0.0062 \cpt 1.1601
0.0022 \cpt 1.1914 0.0002 \cpt 1.2228 0.0002 \cpt 1.2542 0.0022
\cpt 1.2853 0.0062 \cpt 1.3162 0.0120 \cpt 1.3466 0.0199 \cpt
1.3765 0.0296 \cpt 1.4057 0.0411 \cpt 1.4341 0.0545 \cpt 1.4616
0.0696 \cpt 1.4881 0.0865 \cpt 1.5136 0.1049 \cpt 1.5378 0.1249
\cpt 1.5607 0.1464 \cpt 1.5822 0.1693 \cpt 1.6022 0.1935 \cpt
1.6206 0.2190 \cpt 1.6375 0.2455 \cpt 1.6526 0.2730 \cpt 1.6660
0.3014 \cpt 1.6775 0.3306 \cpt 1.6873 0.3605 \cpt 1.6951 0.3909
\cpt 1.7010 0.4218 \cpt 1.7049 0.4529 \cpt 1.7069 0.4843 \cpt
1.7069 0.5157 \cpt 1.7049 0.5471 \cpt 1.7010 0.5782 \cpt 1.6951
0.6091 \cpt 1.6873 0.6395 \cpt 1.6775 0.6694 \cpt 1.6660 0.6986
\cpt 1.6526 0.7270 \cpt 1.6375 0.7545 \cpt 1.6206 0.7810 \cpt
1.6022 0.8065 \cpt 1.5822 0.8307 \cpt 1.5607 0.8536 \cpt 1.5378
0.8751 \cpt 1.5136 0.8951 \cpt 1.4881 0.9135 \cpt 1.4616 0.9304
\cpt 1.4341 0.9455 \cpt 1.4057 0.9589 \cpt 1.3765 0.9704 \cpt
1.3466 0.9801 \cpt 1.3162 0.9880 \cpt 1.2853 0.9938 \cpt 1.2542
0.9978 \cpt 1.2228 0.9998 \cpt 1.1914 0.9998 \cpt 1.1601 0.9978
\cpt 1.1289 0.9938 \cpt 1.0980 0.9880 \cpt 1.0676 0.9801 \cpt
1.0377 0.9704 \cpt 1.0085 0.9589 \cpt 0.9801 0.9455 \cpt 0.9526
0.9304 \cpt 0.9261 0.9135 \cpt 0.9007 0.8951 \cpt 0.8765 0.8751
}
\def\stretchedmuck{\vcenter{%
 \hbox{\box\stretch\unit=50pt
 \point 1.207 0.5 {\rlap{$\to$}\bullet}
 \point 1.207 0.3 c}}\hbox to 85.4pt{\hfil}}

\rightline{\vbox{\hbox{SHEP--91/92--22}\hbox{hep-ph/9207223}}}
\title{Heavy quark symmetry: ideas and applications}

\author{J M Flynn\dag\ and N Isgur\ddag}

\address{\dag\ Physics Department, University of Southampton,
Highfield, Southampton SO9 5NH}

\address{\ddag\ CEBAF, 12000 Jefferson Avenue, Newport News,
Virginia 23606}

\shorttitle{Heavy quark symmetry}


\jnl{\jpg}

\def\monthyear{May 1992}\date

\beginabstract
This report is a combined version of two talks presented by
the authors at the Edinburgh $b$-physics
Workshop, December 1991.  It presents the ideas of heavy quark
symmetry and gives an introduction to some applications.  The
references indicate where to go for more information: they are
not intended to be complete, nor do they necessarily refer to the
original work on any particular subject.
\endabstract

\section{The basic idea: no subtleties}
Consider two hadrons each containing a single heavy quark.  To
make our point, let the heavy quark in the first hadron, $Q_i$,
have a mass of $10\,\rm kg$ and let the other quark $Q_j$, have
a mass of $1\,\rm kg$.
$$
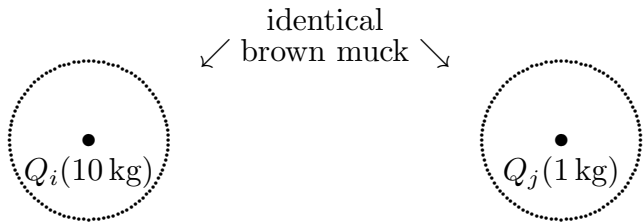
\matrix{
\noalign{\vskip5pt}
 & \swarrow\vbox{\baselineskip10pt
                   \setbox0=\hbox{brown muck}
                   \hbox to\wd0{\hfill identical\hfill}
                   \box0} \searrow & \cr
\noalign{\vskip-5pt}
\muck{Q_i(10\,{\rm kg})} &\qquad \qquad& \muck{Q_j(1\,{\rm
kg})}\cr
\noalign{\vskip5pt}}
$$
As seen in
their rest frames, the hadronic systems that can be built
on each heavy quark out of the light degrees of freedom of QCD
will
be identical:
we are
just looking at how QCD distributes the \lq\lq brown muck''
of light quarks and glue around
a static colour charge.  Since the scale of the interactions of
the brown muck is set by $\Lqcd$ we can say that when
$$
m_{Q_i},\, m_{Q_j} \gg \Lqcd
$$
the heavy quark will in fact be effectively static so that
the light degrees of freedom will be independent of the heavy
flavour.  For $N_h$ flavours of heavy quark there will
be an $\su{N_h}$ symmetry~\jlref\IWstat{{Isgur N and Wise M B
(1989)}
\PL{B 232,}{113.}}
\jlref\IWnonstat{{Isgur N and Wise M B (1990)}\PL{B 237,}{527.}}.
This is {\it not} a model: the
solution of the QCD field equations will be independent of $m_Q$
as $m_Q \to \infty$.  The symmetry is analogous to the isotope
effect in
atomic physics --- the brown muck is independent of the heavy
quark mass just as the electronic structure of an atom is
independent of the number of neutrons in the nucleus.

We can compare this new symmetry to the ordinary $\su3$ flavour
symmetry of the light quarks $u$, $d$, and $s$ which arises since
the light quark masses are {\it small} compared to $\Lqcd$.
In each case one must be careful to apply the symmetry to the
appropriate observables. For example, since the light
flavour symmetry arises from the fact that the light quark masses
are near the chiral limit, and not because they are nearly
degenerate,
it {\it does not} imply that pions and kaons have the same
mass.  In contrast, the
light flavour symmetry {\it does\/} imply that the
baryons built out of light quarks are approximately degenerate.
The pion and kaon are pseudogoldstone bosons whose masses vanish
as the quark masses vanish, whereas the baryon masses have a
finite limiting value.  Similarly, the new symmetry among the
heavy quarks, $c$,
$b$ and $t$ arises because they are much {\it heavier} than
$\Lqcd$.
If the $b$ and $c$ quarks weighed $10\,\rm kg$ and $1\,\rm kg$,
the brown muck distributed around them would obviously look the
same.  The heavy quark symmetry
doesn't say that hadrons containing a single $b$ or $c$ quark
have the same mass (these masses don't
approach a finite value in the heavy quark limit),
but it does say that if you line up the
lowest energy states (at around $10\,\rm kg$ and $1\,\rm kg$
respectively), the spectra will then match.

The strange quark mass is actually not very small compared to
$\Lqcd$, so there are sizeable corrections to the predictions of
light quark symmetry (about 30\%).  Likewise, the deviations from
heavy quark symmetry will be biggest for the charm quark whose
mass is not extremely large compared to $\Lqcd$: one would expect
deviations of about $\Lqcd/m_Q$, or 10\% for charm quarks and 3\%
for $b$ quarks.  Heavy quark symmetry would be best for the top
quark, but it decays so fast via weak interactions that we do not
expect to see top hadrons.

The new symmetry is of an unfamiliar kind.  In the two flavour
example above we saw the symmetry when both heavy quarks were at
rest.  By boosting we can see that the heavy flavour symmetry
will apply between any heavy quarks of the same {\it velocity},
not the same momentum.
$$
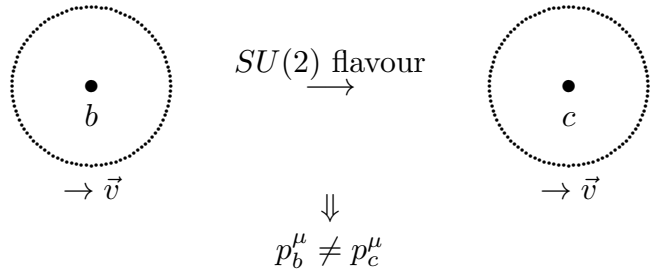
\matrix{
\noalign{\vskip5pt}
\muck{b} & \labelto{\su2\ \hbox{flavour}} & \muck{c}\cr
        \rightarrow \vec{v} & & \rightarrow \vec{v} \cr
\noalign{\vskip-8pt}
        & \Downarrow & \cr
        & p_b^\mu \neq p_c^\mu & \cr
\noalign{\vskip5pt}}
$$
That is, the $\su{N_h}$ maps
$
\ket{H_{Q_i}(\vv,\lambda)} \leftrightarrow
\ket{H_{Q_j}(\vv,\lambda)}
$.
As a result, heavy quark symmetry is a symmetry of certain matrix
elements,
not a symmetry of the $S$-matrix.  For example, the symmetry can
relate form
factors in spacelike regions to those in timelike regions.

In fact there is much {\it more} symmetry
than we have mentioned so far.  Since the heavy quark
spin decouples like $1/m_Q$, just as in atomic physics,
in the heavy quark limit the brown muck doesn't care
about the spin:
$$
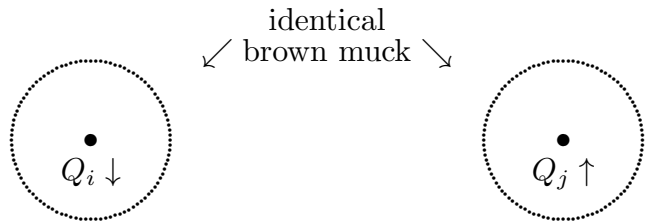
\matrix{
\noalign{\vskip5pt}
 & \swarrow\vbox{\baselineskip10pt
                   \setbox0=\hbox{brown muck}
                   \hbox to\wd0{\hfill identical\hfill}
                   \box0} \searrow & \cr
%                   \hbox to\hsize{\hfill identical \hfill}
%                   \hbox{brown muck}} \searrow & \cr
\noalign{\vskip-5pt}
\muck{Q_i\dn} &\qquad \qquad& \muck{Q_j\up}\cr
\noalign{\vskip5pt}}
$$
The $\su{N_h}$ flavour symmetry becomes an
$\su{2N_h}$ spin-flavour symmetry.  This is reminiscent of the
old $\su6$ quark model, but this time it is exact in the $m_Q \to
\infty$ limit.  The $\su{2N_h}$ is also like Wigner's $\su4$ in
nuclear physics.

\subsection{Spectroscopy}
Since the spin of the heavy quark decouples, we should find
states
occurring in doublets corresponding to the two possible
orientations of the heavy quark spin.  We can classify states
using the heavy quark spin $\vec{S}_Q$ and the remaining \lq\lq
spin'' (combined spin and orbital angular momentum) of the brown
muck, which are both good quantum numbers, and which combine to
make the total spin of the state,
$$
\vec S = \vec{S}_Q + \SL.
$$
Hence we should find degenerate doublets characterised by the
spin $\slt$ of the brown muck, with total spin $\slt \pm \half$
(unless, of course, $\slt=0$).
Of course, heavy quark symmetry can't tell us which $\slt$
quantum
numbers will be associated with which states in the spectrum:
this
is a dynamical issue. In nature we observe (as predicted by the
naive quark
model) that the lowest-lying mesons with $Q \bar q$ quantum
numbers
have $\slt ^{\pi_\ell}={1 \over 2}^-$ ({\it i.e.}, the spin and
parity of an antiquark). Therefore the
degenerate ground state mesons have $J^P=0^-$ and $1^-$
and are the $B$ and
$B^*$ or $D$ and $D^*$ mesons.  Similarly, for baryons we
observe, as expected in the quark model, that the lightest states
with $Qqq$ quantum numbers have
zero light spin ($\slt^{\pi_\ell} = 0+$) giving the $\Lambda_b$
or $\Lambda_c$
baryons, whilst light spin of one ($\slt ^{\pi_\ell} =1^+$) gives
the
degenerate $\Sigma_Q$
and $\Sigma_Q^*$ baryons with $J^P=\half ^+$ or ${3\over2}^+$.
These predictions don't depend on a valence quark approximation,
or the assumption that $\slt$ is dominated by the light quark
spins, but they {\it do\/} depend on the identification of the
$\slt$ multiplets with the physical states.

As mentioned above,
the heavy flavour symmetry tells us that if we line up
the ground states, corresponding to subtracting the mass of the
heavy quark, then the spectra built on different flavours of
heavy quark should look the same.  The splittings are flavour
independent, although the overall scale is not.  This is
illustrated in figure~1.

The four strong transitions between any two pairs of doubly
degenerate states, occurring via the emission of light
hadrons, will be related just by Clebsch-Gordan coefficients.
For example, the following factors relate transitions from $D_1$
and $D^*_2$ to $D\pi$ and $D^*\pi$ states:
$$
\matrix{
 & & &\hbox{\ept\rm relative coefficient}\cr
\noalign{\vskip3pt}
        D_2^* & \to & D\pi\hfill      & \sqrt{2/5}\cr
              &     & [D^*\pi]_S\hfill& 0\cr
              &     & [D^*\pi]_D\hfill& \sqrt{3/5}\cr}
\qquad\matrix{
 & & &\hbox{\ept\rm relative coefficient}\cr
        D_1   & \to & D\pi\hfill      & 0\cr
              &     & [D^*\pi]_S\hfill& 0\cr
              &     & [D^*\pi]_D\hfill& 1\cr}
$$

The double degeneracy of the states is lifted at order $1/m_Q$
where the first spin dependence operates.  The prediction is that
the splitting is $1/m_Q$ times a function at most logarithmic in
$m_Q$.  For the vector and pseudoscalar mesons, if you
approximate $m_Q$ by the average, $(m_V+m_P)/2$, you predict
$m_V^2 - m_P^2$ should be roughly constant.  Experimentally, this
is very well satisfied for $B$, $D$ and even $K$ mesons.
$$
\matrix{
m_{B^*}^2 - m_B^2 & = & 0.55\, {\rm GeV}^2 \cr
m_{D^*}^2 - m_D^2 & = & 0.56\, {\rm GeV}^2 \cr
m_{K^*}^2 - m_K^2 & = & 0.53\, {\rm GeV}^2 \cr}
$$
Initial lattice calculations have been
done~\jlref\bbstar{{Bochicchio M \etal (1992)}\NP{B 372,}{403.}}
to find
the $B$--$B^*$ splitting using heavy quark methods.  However,
even including a perturbative matching
correction~\jlref\jmfbrh{{Flynn J M and Hill B R
(1991)}\PR{D 43,}{173.}} to get from the lattice
result to the continuum value, the result, $(0.19 \pm 0.04 -
0.07)\, {\rm GeV}^2$, is still about one third of the
experimental value.

\subsection{Current matrix elements}
Consider the matrix element of the $b$-number current between a
$\Bbar$ meson of velocity $\vv$ and one of velocity $\vv\,'$.
Compare this to the matrix element of the current $\cbar \g\mu
b$ between a $\Bbar$ of velocity $\vv$ and a $D$ meson of
velocity $\vv\,'$.
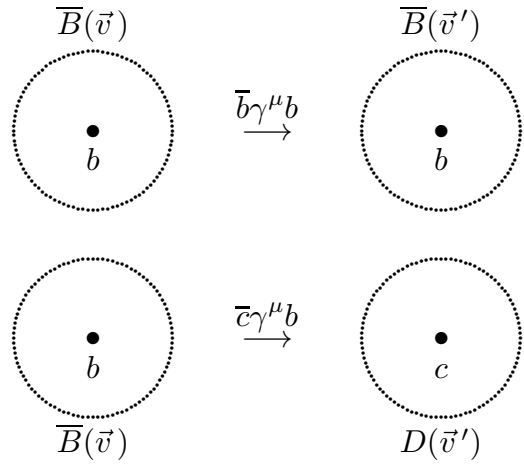
$$
\matrix{
\noalign{\vskip5pt}
\Bbar(\vec{v}\,) & & \Bbar(\vec{v}\,') \cr
  \muck{b} & \labelto{\bbar\g\mu b} & \muck{b} \cr
\noalign{\vskip10pt}
  \muck{b} & \labelto{\cbar\g\mu b} & \muck{c} \cr
 \Bbar(\vec{v}\,) & & D(\vec{v}\,') \cr
\noalign{\vskip5pt}}
$$
The heavy flavour symmetry says the brown muck
doesn't know the difference, since it cares only about the
velocity of the colour sources carried by the heavy quarks.  In
equations:
$$
\eqalign{\bra{\Bbar(\pvp _B)} \bbar \g\mu b \ket{\Bbar(\pv\,)}
&=
  F_B(t_{BB}) (p+p'_B)^\mu \cr
         \bra{D(\pvp _D)} \cbar \g\mu b \ket{\Bbar(\pv\,)} &=
  f_+(t_{DB})(p+p'_D)^\mu + f_-(t_{DB})(p-p'_D)^\mu\cr}
$$
where ${p'_X}^{\mu}=m_X{v'}^{\mu}$ and
$t_{XB}= (p-p'_X\,)^2$.  The symmetry says that $f_\pm$ are
related to $F_B$.  We find, equating coefficients of $v$ and $v'$
and removing the trivial effects of the heavy quark masses from
the normalisation of states,
$$
f_{\pm}(t_{DB})={{m_D \pm m_B} \over {2 {\sqrt{m_Dm_B}}}}
F_B(t_{BB}),
$$
with $t_{DB} = (m_B-m_D)^2 +t_{BB}\, m_D/m_B$.
Furthermore, since $\bbar \g\mu b$ is a symmetry
current, counting $b$-number, the absolute normalisation of $F_B$
is known at $v=v'$ or $t_{BB} = 0$.  Hence we know $f_\pm$ at the
\lq\lq zero-recoil'' point
$t_{DB}=(m_B-m_D)^2$ where a $\Bbar$ at rest decays to a
$D$ at rest.

The spin symmetry of the heavy quark theory lets us say even
more.  We can consider the matrix element of a current $\cbar
\Gamma b$, where $\Gamma$ is {\it any\/} Dirac matrix, between
states where the $b$ and $c$ quarks have {\it any\/} spin, and
relate it to $F_B$.
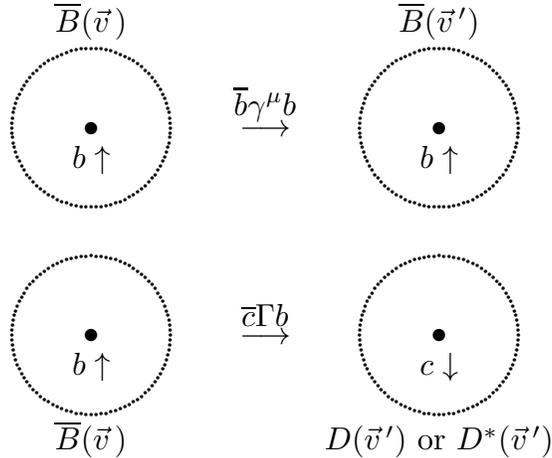
$$
\matrix{
\noalign{\vskip5pt}
\Bbar(\vec{v}\,) & & \Bbar(\vec{v}\,') \cr
  \muck{b\up} & \labelto{\bbar\g\mu b} & \muck{b\up} \cr
\noalign{\vskip10pt}
  \muck{b\up} & \labelto{\cbar\Gamma b} & \muck{c\dn} \cr
 \Bbar(\vec{v}\,) & & \hbox to0pt{\hss$D(\vec{v}\,')$ or
                                  $D^*(\vec{v}\,')$\hss}
 \cr
\noalign{\vskip5pt}}
$$
$F_B$ contains the non-perturbative
information on the response of the brown muck to a change in the
velocity of the colour source from $v$ to $v'$.  The
spin-flavour symmetry tells us that we can use {\it any\/}
current
to kick the heavy quark and change its velocity (and spin and
flavour).

\subsection{One subtlety}
So far we have been economical with the truth.  The preceding
arguments apply in a low energy effective theory with a cutoff
$\mu$, with
$$
\Lqcd \ll \mu \ll m_{Q_j} \leq m_{Q_i}.
$$
Momenta above $m_Q$, however, probe non-static heavy quarks $Q$.
The
results discussed above applied to the current $J$ in the low
energy
theory in which momenta larger than
$\mu$ are cut off.  This current is related to the full current
$j$
according to
$$
j^{ji}_\nu = C_{ji} J^{ji}_\nu + {\rm O}(1/m_Q) + {\rm
O}(\alpha_s/\pi).
$$
That is, since the effective theory and the full theory differ
at high energy there is a calculable perturbative QCD matching
between the full and effective currents.  This matching gives a
correction factor $C_{ji}$ between the two currents, as well as
generating extra operators in the effective theory which match
to the full theory current (the $\alpha_s/\pi$ terms).  There are
also additional corrections of order $1/m_Q$ for heavy quarks
which are not infinitely massive.

If you think of scaling down
from very high energy to low energy the picture looks like this:
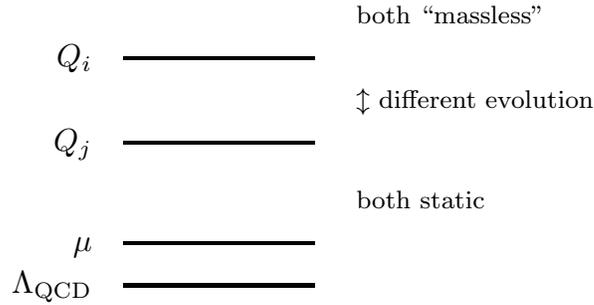
\def\stuff#1{$#1\quad\prop$}
$$\vbox{
\halign{\hfil#&\quad\ept\rm#\hfil\cr
 & both \lq\lq massless'' \cr
\stuff{Q_i} & \cr
 & $\updownarrow$ different evolution \cr
\stuff{Q_j} & \cr
\noalign{\vskip5pt}
 & both static \cr
\stuff{\mu} & \cr
\stuff{\Lqcd} & \cr}}
$$
At scales above the mass of both quarks $Q_i$ and $Q_j$, the full
vector and axial vector
currents are partially conserved and so have zero anomalous
dimension.  The quarks are roughly \lq\lq massless'' and there
is no contribution to $C_{ji}$ in this region.
Once we come below $m_{Q_i}$, however, the $i$ quark is regarded
as heavy whilst the $j$ quark is not, so in this region the
current is not conserved, and $C_{ji}$ is different from
unity~\jlref\VSanomdim{{Voloshin M B and Shifman M A (1987)}{Sov.
J.
Nucl. Phys.}{45,}{292.}} \jlref\PW1{{Politzer H D and Wise M B
(1988)}\PL{B 206,}{681.}} \jlref\PW2{{Politzer H D and Wise M B
(1988)}\PL{B 208,}{504.}}.
Once we move below $m_{Q_j}$, both quarks are heavy.  Again the
current is not conserved unless the quarks have the same
velocity, in which case they are related by the heavy flavour
symmetry.  Hence in the low energy region there is a velocity
dependent contribution to $C_{ji}$ which reduces to~$1$ if $v_j
= v_i$.  We will see this in more detail below.

\section{Heavy Quark Effective Field Theory}
In this section we describe how the ideas of heavy quark
symmetry~[\IWstat] [\IWnonstat] can
be embodied in a low energy heavy quark effective field
theory (HQET)~\jlref\eichtenhill{{Eichten E and Hill B
(1990)}\PL{B 234,}{511.}} \jlref\GeorgiHQET{{Georgi H
(1990)}\PL{B
240,}{447.}}.  This
will allow us to derive Feynman rules for the heavy quarks and
give a recipe for doing calculations.  See the TASI Summer School
lecture notes by Georgi~\bkref\TASI{{Georgi H (1991)}{Lectures
presented at the Theoretical Advanced Study Institute, Boulder
(World Scientific),}{to be published}} for more details.
(Incidentally, several other reviews of heavy quark symmetry
and its applications are available: see~\bkref\louise{{Wise M
B (1991)}{Lectures presented at the Lake Louise Winter Institute,
Caltech preprint}{CALT--68--1721}} \bkref\BGreview{{Grinstein B
(1991)}{Proc. High Energy Phenomenology Workshop, Mexico City,
eds. Huerta R and P\'erez M A, SSCL preprint}{91--17}}
\bkref\IWreview{{Isgur N and Wise M B (1992)}{Proc. Hadron 91,
CEBAF preprint}{TH--92--10}} for example).

First consider a bound state with velocity $v_\mu$, and mass
$M_Q$, which contains a single heavy quark (or antiquark)
together with some brown muck.  The momentum of the bound state
is
$$
P^\mu = M_Q v^\mu.
$$
For a heavy quark we expect the quark mass $m_Q$ to be nearly
equal to the bound state mass, $m_Q \approx M_Q$, with the
difference independent of $m_Q$.  We also expect the
heavy quark to carry nearly all of the momentum of the bound
state, although the brown muck will carry a small momentum
$q^\mu$.  We can write an equation for the quark momentum,
$p^\mu$,
$$
p^\mu = P^\mu - q^\mu = m_Q v^\mu + k^\mu,
$$
where we define the residual momentum, $k^\mu$,
$$
k^\mu = (M_Q-m_Q)v^\mu - q^\mu.
$$
The four velocity of the heavy quark is,
$$
v^\mu_Q = {p^\mu\over m_Q} = v^\mu + {k^\mu\over m_Q}
$$
so that the velocity of the bound state and the quark are the
same in the heavy quark limit.  As $m_Q \to \infty$ the heavy
quark is nearly on-shell and carries nearly all of the bound
state's momentum.

The QCD interactions do not change the heavy quark's velocity at
all.  Any kinks in the trajectory of the heavy quark must be
caused by external, {\it non\/}-QCD, agencies like weak or
electromagnetic interactions.

\subsection{$m_Q \to \infty$ in strong interaction diagrams}
First look at the spinor $u$ for a heavy quark.
Since the final momentum in any strong interaction diagram in
the low energy effective theory will
differ from the initial momentum by an amount much less than
$m_Q$, the heavy quark spinor satisfies
$\slash v u \simeq u$.
Now consider the usual fermion propagator,
$$
{i\over\slash p - m_Q}.
$$
Again, let $p^\mu = m_Q v^\mu + k^\mu$ and look at the limit of
large $m_Q$ to see that,
$$
\eqalign{
{i\over\slash p - m_Q} &= {i(\slash p + m_Q)\over p^2-m_Q^2} =
                          {i(m_Q\slash v + \slash k + m_Q)\over
                           2m_Q v\!\cdot\!k + k^2}\cr
                       &\approx {i\over v\!\cdot\!k}
                          \vproj.\cr}
$$
The $(1+\slash v)/2$ projection operator can always be moved to
a spinor $u$ satisfying $\slash v u = u$ (since, as we will
see below, $\slash v$ commutes with the heavy quark-gluon
vertex), so we replace the
projector by~$1$.  Hence we have a rule for replacing propagators
of heavy quarks according to:
$$
\matrix{
%\noalign{\vskip5pt}
\prop & \quad \longrightarrow \quad & \heavyprop\cr
\noalign{\vskip7pt}
        \displaystyle {i\over\slash p - m_Q} & & \displaystyle
        {i\over v\!\cdot\! k}\cr
%\noalign{\vskip5pt}
}
$$

For the vertex between a heavy quark and a gluon, observe that
it will always occur between propagators or on-shell spinors, so
we can sandwich it between $(1+\slash v)/2$ projectors, and use
$$
\vproj \g\mu \vproj = v^\mu \vproj
$$
to obtain the replacement rule:
\newbox\cycloid
% gluon line
\setbox\cycloid=\hbox{\unit=0.5pt
\cpt 0.0000 6.0000 \cpt 0.4998 5.9896 \cpt 0.9986 5.9585 \cpt
1.4954 5.9067 \cpt 1.9895 5.8343 \cpt 2.4799 5.7412 \cpt 2.9658
5.6277 \cpt 3.4463 5.4938 \cpt 3.9205 5.3396 \cpt 4.3874 5.1653
\cpt 4.8461 4.9709 \cpt 5.2957 4.7567 \cpt 5.7351 4.5228 \cpt
6.1633 4.2695 \cpt 6.5792 3.9969 \cpt 6.9819 3.7055 \cpt 7.3700
3.3954 \cpt 7.7424 3.0671 \cpt 8.0979 2.7209 \cpt 8.4351 2.3573
\cpt 8.7526 1.9769 \cpt 9.0489 1.5802 \cpt 9.3225 1.1679 \cpt
9.5716 0.7409 \cpt 9.7944 0.3001 \cpt 9.9889 -0.1533 \cpt 10.1530
-0.6180 \cpt 10.2844 -1.0924 \cpt 10.3805 -1.5744 \cpt 10.4386
-2.0619 \cpt 10.4555 -2.5516 \cpt 10.4281 -3.0401 \cpt 10.3528
-3.5226 \cpt 10.2257 -3.9932 \cpt 10.0433 -4.4442 \cpt 9.8020
-4.8659 \cpt 9.4994 -5.2457 \cpt 9.1354 -5.5677 \cpt 8.7141
-5.8131 \cpt 8.2464 -5.9621 \cpt 7.7516 -5.9974 \cpt 7.2570
-5.9114 \cpt 6.7910 -5.7096 \cpt 6.3766 -5.4093 \cpt 6.0267
-5.0329 \cpt 5.7457 -4.6020 \cpt 5.5322 -4.1342 \cpt 5.3823
-3.6432 \cpt 5.2912 -3.1389 \cpt 5.2537 -2.6287 \cpt 5.2653
-2.1183 \cpt 5.3216 -1.6118 \cpt 5.4188 -1.1123 \cpt 5.5536
-0.6224 \cpt 5.7227 -0.1439 \cpt 5.9236 0.3214 \cpt 6.1536 0.7724
\cpt 6.4107 1.2080 \cpt 6.6926 1.6274 \cpt 6.9975 2.0298 \cpt
7.3238 2.4147 \cpt 7.6697 2.7814 \cpt 8.0338 3.1297 \cpt 8.4146
3.4591 \cpt 8.8108 3.7693 \cpt 9.2213 4.0600 \cpt 9.6446 4.3310
\cpt 10.0798 4.5820 \cpt 10.5257 4.8130 \cpt 10.9813 5.0238 \cpt
11.4455 5.2142 \cpt 11.9174 5.3841 \cpt 12.3960 5.5335 \cpt
12.8803 5.6622 \cpt 13.3694 5.7702 \cpt 13.8625 5.8576 \cpt
14.3587 5.9241 \cpt 14.8569 5.9699 \cpt 15.3565 5.9949 }
\def\gluon{\vcenter{%
 \hbox{\copy\cycloid}}\hbox to 7.85pt{\hfil}}
$$
\matrix{
\gluon\gluon\gluon\gluon \vcenter{\hrule height 40pt width1pt}
 & \quad\longrightarrow\quad &
\gluon\gluon\gluon\gluon \vcenter{\hbox{\vrule height 40pt
width1pt \kern0.8pt \vrule height40pt width1pt}} \cr
\noalign{\vskip7pt}
\displaystyle -ig \g\mu {\lambda^a\over2} & & \displaystyle -ig
v^\mu {\lambda^a\over2} \cr}
$$
\setbox\cycloid=\hbox{}
where we have again moved the projection operator to act on a
spinor where it gives~$1$.

The new Feynman rules contain no reference to the heavy quark
mass so they show explicitly the symmetry under change of heavy
quark flavour.  Similarly, there are no $\gamma$-matrices in the
Feynman rules, so the heavy quark spin symmetry is also apparent.

In the {\it static\/} theory, where we expand around $v =
(1,0,0,0)$, the propagator and vertex become the nonrelativistic
propagator and charge density coupling respectively,
$$
\eqalign{{i\over v\!\cdot\!k} &\quad \to\quad {i\over k^0} =
 {i\over E-m_Q},\cr
-ig v^\mu {\lambda^a\over2} & \quad\to\quad -ig \delta^{\mu0}
{\lambda^a\over2}.\cr}
$$

\subsection{A systematic expansion}
We would like to have a low energy effective theory which will
allow us to incorporate $\as$ and $1/m_Q$ corrections
systematically~[\GeorgiHQET].
For each velocity $v^{\mu}$, the Feynman rules above
are those arising from a Lagrangian
$$
{\cal L}_v = i \Qbar_v v_\mu D^\mu Q_v
$$
where $D^\mu = \partial^\mu - ig G^\mu_a \lambda^a/2$ is the
colour covariant derivative, and the field $Q_v$ is related to
the ordinary heavy quark field $Q$ by
$$
Q = \exp(-im_Q v\!\cdot\!x) Q_v, \qquad \slash v Q_v = Q_v.
$$
We have done this for the heavy quark only.  There is a similar
procedure to incorporate the heavy antiquark.  The antiquark is
different only in being in a conjugate colour representation.
There is not enough energy to create heavy quark-antiquark pairs,
so the quark and antiquark fields are independent in the HQET
(see~[\TASI] for more details).

The above applies at lowest order in $1/m_Q$.  Corrections
suppressed by inverse powers of the heavy quark mass can be
straightforwardly incorporated.  For example, at dimension five
there are two new operators
$$
\Qbar_v {(iD)^2 - (i v\!\cdot\!D)^2 \over 2 m_Q} Q_v, \qquad
 \Qbar_v {i\sigma^{\mu\nu} D_\mu D_\nu \over 2m_Q} Q_v,
$$
which are the heavy quark kinetic energy and a colour magnetic
moment term, respectively.  The appearance of the
$\sigma^{\mu\nu}$ in the magnetic moment term is the first time
anything has distinguished the heavy quark spins.  Hence, the
magnetic moment operator will be responsible for splitting the
vector and pseudoscalar heavy-quark meson masses at order
$1/m_Q$.

The underlying spin and flavour symmetries are easy to see in the
effective lagrangian,
$$
{\cal L} = \sum_{\vec{v}} \sum_{j=1}^{N_h} i\Qbar_{j_v}
v\!\cdot\!D  Q_{j_v}.
$$
For each $\vec{v}$ we can rotate any spin component of
any flavour of heavy quark into another, giving the $\su{2N_h}$
spin-flavour symmetry.  Lorentz transformations mix up the
different velocities of heavy quark.  The overall symmetry has
been christened \lq\lq $\su{2N_h}^\infty \otimes
\hbox{Lorentz}$'' by Georgi.

\subsection{Relation to QCD}
By construction, the heavy quark effective theory (HQET)
described
in the two steps above reproduces the low energy behaviour of
QCD: we build the heavy theory demanding that it give the same
$S$-matrix elements as QCD.  This means that the naive heavy
quark limit must be corrected for the effects of high energy
virtual processes.  For example, the weak flavour changing $b$
to $c$ current is corrected at order $\as$ in both QCD
and the HQET.  The difference between these corrections tells us
how we must modify the coefficient of the HQET current, as well
as possibly introducing new structures in the HQET current, so
that the HQET reproduces the physics.

We illustrate for the case of the current $\bbar \g\mu c$.  The
$\g\mu$ in the QCD current is replaced in the HQET current
by~\jlref\neubert{{Neubert M (1992)}\NP{B 371,}{149.}}
$$
\g\mu \longrightarrow \hqgamma^\mu = \left(1+C_0
 {\as\over\pi}\right) \g\mu +
 {\as\over\pi} \sum_i C_i \Gamma^\mu_i
$$
so there is a new strength for the naively matched $\g\mu$
together with new structures $\Gamma^{\mu}_i$ (such as $v^\mu$
and
$v'{}^\mu$, where $v$
and $v'$ are the heavy quark velocities).  We say that we \lq\lq
match the low energy approximation to the full theory''.
Diagrammatically, the two sets of diagrams shown in figure 2 are
matched to determine $\hqgamma$ to order $\as$ (the figure
illustrates the case where both quarks are treated as heavy in
the effective theory).

If $m_b \gg m_c \gg \mu$, the new coefficient of the naive
$\g\mu$ turns out to have the form shown above with
$$
C_0 = \ln {m_b\over m_c} - {4\over3}[w\, r(w) - 1]
      \ln{m_c\over\mu}
$$
where $w = v\!\cdot\!v'$ and
$$
r(w) = {\ln (w + \sqrt{w^2-1}\,)\over \sqrt{w^2-1}}.
$$
The leading logs can be summed by the renormalisation group with
the result~\jlref\cusp{{Korchemsky G P and Radyushkin A V
(1987)}\NP{B 283,}{342.}} \jlref\FGGWanomdim{{Falk A F, Georgi
H,
Grinstein B and Wise M B (1990)}\NP{B 343,}1.},
$$
\hqgamma^\mu = C_{cb} \g\mu + {\as\over\pi}
   \sum_i C_i \Gamma^\mu_i
$$
with
$$
C_{cb} = \left[ \as(m_b)\over\as(m_c) \right]^{-6/25}
         \left[ \as(m_c)\over\as(\mu) \right]^{8[w\,r(w)-1]/27}.
$$
The $\mu$-dependence in $C_{cb}$ is cancelled by the
$\mu$-dependence of the non-perturbative function describing the
brown muck transition from $v$ to $v'$.
The two factors in $C_{cb}$ correspond to the renormalisation of
the current
between $m_b$ and $m_c$ and then between $m_c$ and $\mu$,
respectively.  As advertised earlier, we see explicitly that the
second factor is~$1$ when $w=1$, in which case the heavy flavour
symmetry relates the heavy $b$ and $c$ quarks, so that the
current is conserved and has no anomalous dimension.  Strictly
speaking, if we perform the matching at order $\as$ we should use
the two loop anomalous dimension for the renormalisation group
scaling.  Then we match once at the $b$ quark scale, where the
$b$ quark becomes heavy, scale between $m_b$ and $m_c$ and match
again at $m_c$ where the $c$ quark becomes heavy.  The
anomalous dimension calculations have now been taken to two
loops~\jlref\broadhurst{{Broadhurst D J and Grozin A G
(1991)}\PL{B 267,}{105.}} \jlref\jimusolf{{Ji X and
Musolf M J (1991)}\PL{B 257,}{409.}}, so the matching and scaling
can be done.

\section{Some Applications}
Whenever a symmetry can be identified it gives us calculational
power.  By
relating various matrix elements, and fixing the
absolute normalisation of some, heavy
quark symmetry enhances our
predictive ability, allowing us in some cases to finesse
the difficulties of
understanding hadronic structure.  This is analogous to pion,
kaon,
and eta
physics where the effective chiral lagrangian can be used to
extract systematically the consequences of the pattern of chiral
symmetry breaking.  For heavy quarks we have identified a new
symmetry and have developed the heavy quark effective theory to
calculate its predictions.

Heavy quark ideas also help us to model $b$ quarks in lattice
calculations, giving access to the actual values of further
matrix elements.  The problem with putting $b$ quarks on the
lattice by conventional methods is that their Compton wavelength
is smaller than the lattice spacing, and the $b$ quarks \lq\lq
fall through''.  Heavy quark symmetry allows us to extract the
$b$ mass dependence, leaving an effective theory without such a
large mass scale, which can be modelled on present day lattices.
These ideas were discussed elsewhere at this workshop, and we
refer readers to the lattice group's contributions.
Here we will concentrate on one
area of great potential for heavy quark ideas:
constraining the CKM matrix using $b$-physics.

\subsection{Determining $V_{cb}$}
One of the first applications of the ideas of heavy quark
symmetry was to
semileptonic $\Bbar$ meson decays and the extraction of the $b$
to
$c$ mixing angle~\jlref\volshif{{Voloshin M B and Shifman M A
(1988)}{Sov. J. Nucl. Phys.}{47,}{511.}}
[\IWstat] [\IWnonstat].

$V_{cb}$ can be determined from semileptonic decays of $\Bbar$
mesons to $D$ and $D^*$ mesons.  There are altogether six form
factors in the two decays,
$$
\eqalign{
\bra{D(p')} V^\mu \ket{\Bbar(p)} &=
   f_+(t_{DB})(p+p')^\mu + f_-(t_{DB})(p-p')^\mu, \cr
\bra{D^*(p'\!,\epsilon)} A^\mu \ket{\Bbar(p)} &=
   f(t_{DB})\epsilon^{*\mu} + a_+(t_{DB}) \epsilon^*\!\!\cdot\!p
   (p+p')^\mu + a_-(t_{DB}) \epsilon^*\!\!\cdot\!p (p-p')^\mu,
   \cr
\bra{D^*(p'\!,\epsilon)} V^\mu \ket{\Bbar(p)} &=
   i g(t_{DB}) \epsilon^{\mu\nu\lambda\sigma} \epsilon^*_\nu
   (p+p')_\lambda (p-p')_\sigma. \cr}
$$
In these equations, $V^\mu = \cbar \g\mu b$, $A^\mu = \cbar \g\mu
\gamma_5 b$ and $t_{DB} = (p-p')^2$.  In section 1.2 we saw how
the
heavy flavour symmetry related the vector current matrix element
to that of the $b$-number current.  Now the spin symmetry relates
{\it all\/} the $\Bbar \to D$ and $\Bbar \to D^*$ matrix
elements, so they can each be
expressed in terms of {\it one\/} universal function $\xi(w)$,
where $w
= v\!\cdot\!v'$.  This function is the {\it same\/} for any heavy
quark transition, $Q_i \to Q_j$ with the same brown muck.  It
describes the response of the brown muck to the change in
velocity of the colour source from $v$ to $v'$.  Furthermore,
when $v=v'$ there is a flavour symmetry between the two heavy
quarks.  Then the current causing the transition is a symmetry
current so the normalisation of $\xi(w)$ is fixed at the point
$w=1$ which is maximum $t_{DB}$.  We can take the normalisation
to be $\xi(1) =1$.  For
$\Bbar$ to $D^{(*)}$ decays this point is the \lq\lq zero
recoil'' point where a $\Bbar$ at rest decays to a $D$ or $D^*$
at rest.

The relations of the form factors to $\xi$ come out as follows:
$$
\eqalign{
f_\pm &= C_{cb}\, {m_D\pm m_B\over2\sqrt{m_Bm_D}}\, \xi(w),
\cr
g = a_+ = -a_- &= C_{cb}\, {1\over2\sqrt{m_Bm_D}}\, \xi(w), \cr
f &= C_{cb} (w+1) \sqrt{m_Bm_D}\, \xi(w). \cr}
$$
The factor $C_{cb}$ is the perturbatively calculable correction
to the strength of the current in the HQET that we described
above.  For details see~[\FGGWanomdim].

If experimental measurements can be reliably extrapolated to the
zero recoil point then we can determine the CKM matrix element
$V_{cb}$.  Alternatively, we can actually calculate $\xi(w)$ on
the lattice, by looking at the $\Bbar$ to $D^{(*)}$ matrix
element as a function of the velocity transfer between the heavy
quarks.

\subsection{Determining $V_{ub}$}
The idea
behind using heavy quark symmetry
to extract the $b \to u$ mixing angle is illustrated
by the following picture~\jlref\IWVub{{Isgur N and Wise M B
(1990)}\PR{D 42,}{2388.}}:
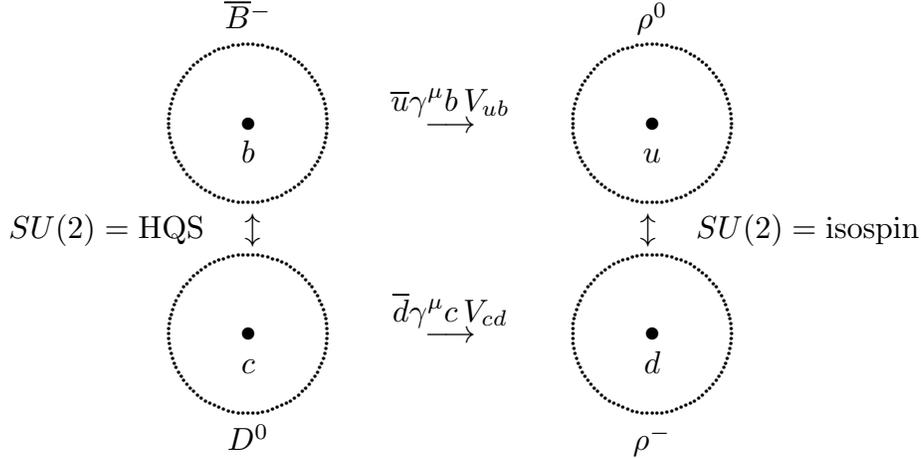
$$
\matrix{\noalign{\vskip5pt}
\Bbar^- & & \rho^0\cr
\muck{b} & \labelto{\overline{u}\g\mu b\,V_{ub}} & \muck{u} \cr
\llap{$\su2  = {\rm HQS}\quad$} \updownarrow & & \updownarrow
                     \rlap{$\quad\su2 = {\rm isospin}$} \cr
\muck{c} & \labelto{\overline{d}\g\mu c\,V_{cd}} & \muck{d} \cr
D^0 & & \rho^- \cr\noalign{\vskip5pt}}
$$
The heavy quark symmetry relates $\Bbar \to \rho^0$
to the unphysical $D \to
\rho^0$ matrix elements.  However, light quark isospin relates
$\rho^0$ and $\rho^-$, and the decay $D^0 \to \rho^-$ is
determined by the known CKM matrix element $V_{cd}$.
This means that we can close the loop in the diagram above to fix
$V_{ub}$ from the $\Bbar^- \to \rho^0$ decay.  Note that the two
$\rho$ states on the right hand side above are \lq \lq pure brown
muck'':
the heavy quark symmetry relates matrix elements for $b$ and $c$
quarks surrounded by the same brown muck to go to the same pure
brown muck states.

The matrix elements of the vector and axial vector weak currents
between $\Bbar$ or $D$ and $\pi$ or $\rho$ involve six form
factors (defined like those in $\Bbar \to D^{(*)}$ decay).  Since
the heavy quark symmetry relates states of the same velocity, one
difficulty is that the allowed range of momentum transfer in the
$D$ decays does not allow us to cover the whole range for the
$\Bbar$ decays of interest.
In its purest form, the above extraction must therefore be done
in the region of the $\Bbar \rightarrow \rho$ Dalitz plot where
the $\rho$
momenta do not exceed those available in the $D \rightarrow \rho$
Dalitz plot. Another potential difficulty may be illustrated by
the example
of the $f_+$ form factor.
The value of this form factor
in $\Bbar$ decay is determined by {\it both} $f_+$ and $f_-$ for
$D$ decay, but the $f_-$ contribution to the $D$ decay rate is
suppressed by a factor of $(m_\ell/m_D)^2$, where $m_\ell$ is the
mass of a charged lepton. Since the decay to the $\tau$
lepton is not kinematically allowed, $f_-$ will be very difficult
to obtain from $D$ decay. Fortunately, one can show that in the
heavy
quark limit there are enough relationships amongst heavy-to-light
form factors to overcome this lack of data [\IWVub]. In this
case,
$f_-=-f_+$ up to corrections of order $\Lambda_{QCD}/m_c$.

\subsection{Rare $B$ decays}
Rare $B$ decays are expected to be a good probe of new physics,
but if we are to see new physics we had better know the standard
model expectation first.  HQET can help us~[\IWVub] by relating
the matrix
elements of interest, such as in $\Bbar \to \overline{K} e^+ e^-
$, to more easily measured processes like $D \to K^- e^+ \nu_e$.
In fact for these examples we need the matrix element of the
vector current between the heavy and light pseudoscalar meson
states.  Using light quark flavour symmetry, this is the same
problem as looking at $\Bbar \to \pi$ or $D\to \pi$ matrix
elements considered in section~3.2.

The rare decays $\Bbar \to K^* \gamma$ and $\Bbar \to K^* e^+ e^-
$ have contributions from a transition magnetic moment operator,
$$
\overline{s}_L \sigma^{\mu\nu} b_R.
$$
We can use the heavy quark spin symmetry to relate the matrix
elements of this operator to those of a current.  Set $\mu=0$ and
$\nu=i$ and observe that,
$$
\sigma^{0i} \propto [\g0,\g i] = -2 \g i \g0.
$$
In the rest frame of the heavy $b$ quark, $\g0 b = b$, so, in the
rest frame we find,
$$
\overline{s}_L \sigma^{\mu\nu} b \quad
\buildrel \hbox{related to} \over \longrightarrow \quad
\overline{s}_L \g i b_L.
$$
Furthermore, the heavy flavour symmetry allows us to relate the
current to $\overline{s}_L \g i c_L$, so that $\Bbar \to K^*
\gamma$ is related to $D^0 \to K^- e^+ \nu_e$.  One problem here
is that in $\Bbar \to K^* \gamma$, the $K^*$ has a fixed momentum
outside the kinematic range of the corresponding kaon in the
semileptonic $D$ decay (this was discussed at the workshop by
A. Ali).

\subsection{Heavy baryon weak decays}
States in which the brown muck has spin $\slt = 0$ form
spin-$1/2$ baryons.  The heavy quark spin symmetry relates
up and down spin states of the baryon, so it will relate baryon
form factors among themselves.

    As mentioned in section~1.1, the
lowest-lying $Qqq$ state containing a single heavy quark
is expected to be a $\Lambda_Q$ with $\slt ^{\pi_\ell}=0^+$.
The simplest example is the $\Lambda$ where a
strange quark is bound to an isospin zero
$\slt ^{\pi_\ell}=0^+$ light quark state.
The $\Lambda_c$ has been observed with a mass of $2285\,\rm MeV$,
but the corresponding $\Lambda_b$ is yet to be
confirmed~\jlref\Lamb1{{Basile M \etal (1981)}{\NC\
Lett.}{31,}{97.}}
\jlref\Lamb2{{Arenton M W \etal (1986)}\NP{B 274,}{707.}}.  Heavy
quark symmetry tells us that the mass splitting of the
pseudoscalar meson and the baryon is independent of the heavy
quark mass in leading order in the HQET, so we expect
$m_{\Lambda_b} = m_B + m_{\Lambda_c} - m_D$.

The rate for the semileptonic decay $\Lambda_b \to \Lambda_c e
\overline{\nu}$ is given in terms of six form factors:
$$
\eqalign{
\bra{\Lambda_{Q_j}(\smash{p',s'})} V^\mu \ket{\Lambda_{Q_i}(p,s)}
 &= \overline{u}^{s'}\!(p') \left( F_1^{ji} \g\mu + F_2^{ji}
         v^\mu + F_3^{ji} v'^\mu \right) u^s(p) \cr
\bra{\Lambda_{Q_j}(\smash{p',s'})} A^\mu \ket{\Lambda_{Q_i}(p,s)}
 &= \overline{u}^{s'}\!(p') \left( G_1^{ji} \g\mu \g5 + G_2^{ji}
         v^\mu \g5 + G_3^{ji} v'^\mu \g5 \right) u^s(p) \cr}
$$
where $p = m_{Q_i} v$ and $p' = m_{Q_j} v'$.  Heavy quark
symmetry implies that,
$$
F_1^{ji} = G_1^{ji} = C_{ji}\, \eta(w)
$$
where $C_{ji}$ is the same renormalisation factor we discussed
earlier, arising from matching the HQET to QCD: it depends only
on the heavy quarks.  The function $\eta(w)$ is a universal
(brown muck dependent) function of the velocity transfer,
$w=v\!\cdot\!v'$.  The remaining form factors are zero in the
heavy quark limit.

Heavy quark symmetry makes the same prediction for decays of
$Qsu$ and $Qsd$ baryons, $\Xi_b \to \Xi_c \enu$, and similar
predictions for the decay of $Qss$ baryons,
$$
\Omega_b \to \Omega_c \enu,\ \Omega_c^* \enu.
$$
One can also prove that the following decays are forbidden in the
heavy quark limit ($\Sigma_Q$ is a $Quu$ or $Qdd$ baryon):
$$
\eqalign{\Lambda_b &\to \Sigma_c \enu \cr
\Lambda_b &\to \Sigma_c^* \enu \cr} \qquad\hbox{and}\qquad
\eqalign{\Xi_b \to \Xi'_c \enu \cr \Xi_b \to \Xi_c^* \enu \cr}
$$
In each case we find a common QCD correction to the matrix
element, determined by the renormalisation of the current which
changes $b$ into $c$, together with a function of $w = v\!\cdot\!
v'$ which contains the response of the brown
muck~\jlref\IWbaryons{{Isgur N and Wise M B (1991)}\NP{B
348,}{276.}}
\jlref\Georgibaryons{{Georgi H (1991)}\NP{B 348,}{293.}}.

Heavy quark symmetry can be applied to some purely hadronic
decays.  For example, $\Lambda_b \to \Lambda_c D_s$ can be
related to $\Lambda_b \to \Lambda_c D^*_s$, since the underlying
process, $b \to c \cbar s$ involves three heavy quarks (the $c$
and $\cbar$ are independent in the HQET)
\bkref\hadronic{{Grinstein B, Kilian W, Mannel T and Wise M B
(1991)}{Harvard preprint}{HUTP 91/A005}}.

\subsection{Factorisation}
Attempts have long been made to justify factorisation in two body
decays of pseudoscalar mesons.  Factorisation means that the
matrix element for $\Bbar \to D \pi$, for example, can be
separated
as,
$$
\bra{D\pi} \overline{d} \gamma_\mu u \cbar \g\mu b \ket{\Bbar}
\approx \bra{D} \cbar \g\mu b \ket\Bbar
        \bra{\pi} \overline{d} \gamma_\mu u \ket0.
$$
This cannot really be true since the two sides have different
renormalisation point dependence.  However, with some extensions
of HQET ideas it has been possible to
prove~\jlref\factorisation{{Dugan M and Grinstein B (1991)}\PL{B
255,}{583.}} that,
$$
\Gamma(\Bbar \to D\pi) = 6\pi^2 f_\pi^2 A^2
 \left. \d\Gamma(\Bbar \to D\enu) \over \d m^2_{\enu}
 \right|_{m^2_{\enu} = m^2_\pi}
$$
with corrections
of order $\Lqcd/m_c$. Here
$A \approx 1.15$ is a renormalisation group matching
factor.  This relation agrees well with experiment.

The same method suggests that factorisation need {\it not} hold
in
$\Bbar \to \pi\pi$ or $\Bbar \to D \overline{D}$, but it does
suggest a test: factorisation should hold for $\Bbar \to D
\pi\pi$ when the two final state pions are collinear.

Factorisation and heavy quark symmetry give absolute predictions
for ratios of decay rates, for example, $\Gamma(\Bbar \to
D\pi)/\Gamma(\Bbar \to D^*\pi)$.  The order $\as(m_b)/\pi$
corrections
(both factorisable and non factorisable) to this result are
small~\jlref\PWfactor{{Politzer H D and Wise M B (1991)}\PL{B
257,}{399.}}, although their precise value is not well determined
because of cancellations.

\subsection{A Bjorken sum-rule for $\xi(w)$}
Consider the following picture in which a $b$ quark with velocity
$v$ surrounded by brown muck is kicked by a $\cbar \g\mu b$
current, giving a $c$ quark moving at velocity $v'$.  The brown
muck must rearrange itself and reform with some surrounding the
moving $c$ quark, and some possibly left behind.  Clearly we have
a set of possible outcomes where the $c$ quark has all possible
brown muck configurations reachable from the initial one, each
outcome having an associated probability.  This will give us a
sum rule~\bkref\BJ{{Bjorken J D (1990)}{Invited talk at Les
Rencontres de Physique de la Vallee d'Aoste, La Thuile, SLAC
preprint}{SLAC--PUB--5278}}
\jlref\IWsumrule{{Isgur N and Wise M B (1991)}\PR{D 43,}{819.}}.
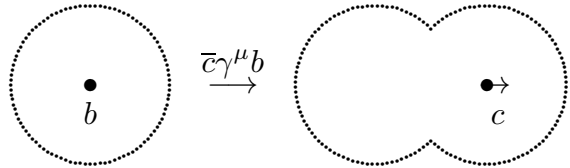
$$
\muck{b} \labelto{\cbar\g\mu b} \stretchedmuck
$$

Since the brown muck cannot change the heavy quark velocity in
the heavy quark limit (the \lq\lq velocity superselection rule'')
one can obtain the following sum rule:
$$
\hbox{Rate}[ b(\vec{v}\,) \to c(\vec{v}\,{}')] =
\sum_{X_c} \hbox{Rate}[ \Bbar(\vec{v}\,) \to
X_c(\vec{v}\,{}')].
$$
The inclusive rate for a $b$ quark of velocity $\vec v$ to go to
a $c$ quark of velocity $\vec{v}\,{}'$ is obtained by summing
over all possible states containing a $c$ quark of velocity
$\vec{v}\,{}'$ which can be reached from a $\Bbar$ containing a
$b$ quark of velocity $\vec v$.  That is, the heavy quark is
undisturbed by the \lq\lq splash'' of the brown muck.
Explicitly:
$$
\eqalign{
1 &= {w+1\over2}\, |\xi(w)|^2 \cr
  &\phantom{=} + \half(w-1)^2(w+1) \sum_{n=1}^{n_{\rm max}(\mu)}
                 |\xi^{(n)}(w)|^2 \cr
  &\phantom{=} + 2(w-1) \sum_{m=1}^{m_{\rm max}(\mu)}
                 |\tau^{(m)}_{1/2}(w)|^2 \cr
  &\phantom{=} + (w-1)(w+1)^2 \sum_{p=1}^{p_{\rm max}(\mu)}
                 |\tau^{(p)}_{3/2}(w)|^2 \cr
  &\phantom{=} + \cdots \cr}
$$
The successive lines on the right hand side refer first to the
sum over $D$ and $D^*$ followed by the sums over other states
with light spin $\slt^{\pi_\ell} = \half^-$, then
$\slt^{\pi_\ell} = \half^+$ and $\slt^{\pi_\ell} = {3\over2}^+$
and so on.  The heavy quark symmetry makes the RHS doable.  For
example, each $\slt^{\pi_\ell} = {3\over2}^+$ multiplet has eight
form factors all related to a single function $\tau_{3/2}$.

The interpretation is that as rate disappears from the \lq\lq
elastic'' channel ($D^{(*)}$) as you move away from $w=1$, it
appears in the excited states.  This is analogous to the
Cabibbo--Radicati sum rule for the proton form factor.  If you
write the universal function $\xi$ as
$$
\xi(w) = 1 - \rho^2(w-1),
$$
where $\rho$ is the slope or \lq\lq charge radius'', you can
prove that
$$
\rho^2 = {1\over4} + \sum \left| \tau^{(m)}_{1/2}(1)\right|^2 +
         2 \sum \left| \tau^{(p)}_{3/2}(1) \right|^2,
$$
{\it i.e.}, the
compensation is only by the $\slt^{\pi_\ell} =
\half^+,{3\over2}^+$ states, and $\rho^2 \geq {1\over4}$.  In the
harmonic oscillator
quark model, the compensating states are the lowest excitations
of the brown muck, obtained by combining a light quark of
spin-${1\over 2}$ with an orbital angular momentum of 1.  The
$\slt^{\pi_\ell} = {3\over2}^+$ states are known and
$\rho^2$ is now measured.  This sum rule should soon give an
interesting constraint on theory/experiment.

There is an analogous sum rule for the
$\Lambda_b$~\jlref\Lbsumrule{{Isgur N, Wise M B and Youssefmir
M (1991)}\PL{B 254,}{215.}}:
$$
\eta(w) = 1 - \rho_\Lambda^2(w-1) + \cdots
$$
where
$$
\rho_\Lambda^2 = 0 + \sum \left| \sigma^{(n)}(1) \right|^2.
$$
This time, the one-quarter is replaced by a zero and the sum is
over the states with $\slt^{\pi_\ell} = 1^-$, which are again the
lowest excitations in the quark model.

\subsection{$\Lqcd/m_Q$ corrections vanish at $w =1$}
Heavy quark symmetry makes many predictions in the $m_Q \to
\infty$ limit.  However, since the quarks to which we wish to
apply these
ideas are not really infinitely massive, we must ask about
$\Lqcd/m_Q$ corrections.  Fortunately, the effective field theory
organises these correction effects, so there is some predictive
power from the symmetry even when it is broken~\jlref\luke{{Luke
M E (1990)}\PL{B 252,}{447.}}.  This is reminiscent of the
Gell-Mann--Okubo and Coleman--Glashow formulas.

The most surprising case so far is for the form factors in the
matrix elements for the heavy baryon decay $\Lambda_b \to
\Lambda_c$~\jlref\GGW{{Georgi H, Grinstein B and Wise M B
(1990)}\PL{B 252,}{456.}}.  Just one new constant, $\Delta m$ is
required, and
the universal function $\eta(w)$ is still normalised, $\eta(1)
= 1$.  The situation is summarised as follows:
$$\vbox{\baselineskip=14pt
\halign{$#$\qquad\hfil&$#$\quad\hfil&$#$\hfil\cr
 & \hbox{\ept\rm leading order}
 & \hbox{\ept\rm with $\Lambda/m_Q$ corrections} \cr
\noalign{\vskip3pt}
F_1 & \eta(w) & \eta(w) (1+\overline{\Delta}) \cr
F_2 & 0       & -\eta(w) \overline{\Delta} \cr
F_3 & 0       & 0 \cr
G_1 & \eta(w) & \eta(w) \cr
G_2 & 0       & -\eta(w) \overline{\Delta} \cr
G_3 & 0       & 0 \cr}}
$$
The $1/m_Q$ corrections are given in terms of $\overline{\Delta}
= \Delta m /[m_c(1+w)]$.
Perturbative QCD corrections will add a multiplicative correction
factor $C_{cb}$ to all terms, as we discussed above.  Then the
result is good up to corrections of orders, $(\Delta m/m_c)^2$,
$\Delta m/m_b$ (only the $c$-quark was given a finite mass) and
$\as (\Delta m/m_c)$.

The fact that the form factor $G_1$ retains a known normalisation
at zero recoil $w=1$ offers the possibility of extracting
$V_{cb}$ with reasonably small uncertainties.  In fact, at zero
recoil there are {\it no\/} $1/m_Q$ corrections to the matrix
elements of the vector and axial vector
currents~\jlref\boydbrahm{{Boyd C G and Brahm D E (1991)}\PL{B
257,}{393.}}.
To understand
why, recall that the charges associated with $V^\mu$ and $A^\mu$
are symmetry generators when $v=v'$, and $b$ quarks can be
rotated into $c$ quarks of the same velocity.  More explicitly,
write the ground state baryon for finite $m_Q$ as a perturbative
sum over states in the $m_Q \to \infty$ limit:
$$
\ket{\smash{\psi_{m_Q}^0}} = \ket{\smash{\psi_\infty^0}} +
  \sum_{n\neq0}\ket{\psi_\infty^n}
  {\bra{\psi_\infty^n} {\rm O}(\Delta m/m_Q)
  \ket{\smash{\psi_\infty^0}} \over (E_n - E_0)}.
$$
Since the $m_Q \to \infty$ states are eigenstates of the charges,
we have $\bra{\smash{\psi_\infty^0}} Q \ket{\psi_\infty^n} =
\delta^{n0}$, so that,
$$
\bra{\psi_{m_2}^0} Q \ket{\psi_{m_1}^0} = 1 + {\rm O}(\Delta
  m/m_Q)^2.
$$

For semileptonic $\Bbar \to D^{(*)}$ decays things are not so
simple.  The new constant $\Delta m$ enters and only two of the
six form
factors are unaffected by $1/m_Q$ corrections.  In particular,
the form factor proportional to $v^\mu + v'{}^\mu$ in the vector
current matrix element is unaffected.  However, experiments
measure a form factor proportional to $p^\mu + p'{}^\mu = m_b
v^\mu + m_c v'{}^\mu$ (the form factor accompanying $p^\mu -
p'{}^\mu$ picks up a factor of the lepton mass when contracted
with the leptonic part of the matrix element for the decay),
which contains an admixture of a corrected form factor.
Fortunately, at zero recoil, the $\Bbar \to D^*$ matrix
element depends solely on an uncorrected form factor, so this may
be the best way to extract $V_{cb}$ from semileptonic $\Bbar$
decays.  Initial analysis by Neubert
gives~\jlref\NeubertVcb{{Neubert M (1991)}\PL{B 264,}{455.}},
$$
|V_{cb}| \left(\tau_B \over 1.18\,{\rm ps} \right) =
   0.045 \pm 0.007
$$
which is slightly more precise than values extracted using model
dependent analyses, and is {\it model independent}.

\section{Status and prospects}
The number of papers on heavy quark symmetry produced in the last
two years is in the hundreds.  New papers on this subject appear
nearly every day.
Corrections for finite heavy quark
masses and for perturbative QCD matching have been classified and
calculated, and some phenomenology done.  The absence of $1/m_Q$
corrections at $v=v'$ is possibly the most significant recent
development.

%HQET ideas have been applied to justify the often-used formula
%giving the inclusive width for semileptonic $\Bbar$ decay by the
%underlying quark decay,
%$$
%\sum_{X_c} \Gamma(\Bbar \to X_c \enu) \approx
%  \Gamma(b\to c \enu)
%$$
%In $e^+e^-$ annihilation, heavy quark flavour symmetry relates
%$\sigma(e^+e^- \to \Bbar B)$ to $\sigma(e^+e^- \to \overline{D}
%D)$ at the appropriately related centre of mass energies.  The
%spin symmetry also relates the production rates of $\Bbar B$,
%$\Bbar B^*$ and $\Bbar^* B^*$, confirming earlier predictions.

    It now appears that it
may be useful to think of hadrons containing a single heavy quark
as the \lq\lq
hydrogen atoms of QCD''.  There are many advantages to this
limit:
relativistic effects are simplified and the heavy quark acts as
a pointlike probe of the light \lq\lq constituent quarks''.  The
symmetries and rigorous results of
the heavy quark limit can be used for
consistency checks (in the form of ``boundary conditions") on
models.

   In a more practical vein,
new data from beauty and charm factories should allow us both to
test
heavy quark symmetry and obtain tighter limits on standard model
parameters.  Heavy quark ideas applied to lattice calculations
may allow the theoretical prediction of strong interaction matrix
elements which were unavailable before.

Ironically, heavy quarks may prove to be an essential tool
in finally helping us to understand the nature
of the brown muck of QCD.

\listrefs

\vfill\eject

% reclaim some memory
\setbox\circle=\hbox{}
\setbox\stretch=\hbox{}

\def\stackup#1#2{\vcenter{\hbox{$#1$\kern3pt}
\hbox{$#2$\kern3pt}}}
\def\st{\vrule height7.5pt depth0.8pt width0pt}
$$\vbox{\offinterlineskip\eightpoint
\halign{$#$\hfill&\hfill$#$\hfill&
\quad$#$\hfill&\hfill$#$\hfill&\quad#\hfill\cr
 & \vdots & & \vdots & \cr
 & \heavyprop & & \heavyprop & $\leftarrow$
\vbox{\hbox{\ept\rm\st levels characterised by $\slt$}
\hbox{\ept\rm\st total spin: $s = \slt \pm 1/2$}} \cr
\noalign{\vskip10pt}
 & \vcenter{\propline\kern0.8pt\propline\kern20pt
\propline\kern0.8pt\propline} & &
\vcenter{\offinterlineskip
\propline\kern0.8pt\propline\kern20pt\propline
\hbox{\vbox to 0pt{\vss\hbox{\kern14.8pt\vrule height23pt
      \kern9.6pt\vrule height21.5pt
      \kern9.6pt\vrule height23pt
      \kern9.6pt\vrule height21.5pt}}}
\hbox{\kern34.8pt\vrule height0.8pt
      \kern9.6pt\vrule height0.8pt}
\propline} &
$\leftarrow$  \vbox{%
\hbox{\ept\rm\st strong transitions related by}
\hbox{\ept\rm\st Clebsch--Gordan coefficients}} \cr
\noalign{\vskip5pt}
\stackup{D_2^*(2460)}{D_1(2420)} & \heavyprop &
 & \heavyprop & \cr
\noalign{\vskip15pt}
\stackup{D^*(2010)}{D(1870)} & \heavyprop &
\stackup{B^*(5330)}{B(5280)} & \heavyprop & $\leftarrow$
\hbox{\ept\rm line up groundstates} \cr
\noalign{\vskip8pt}
 & D \hbox{\ept\rm\ states} &
 & B \hbox{\ept\rm\ states} & \cr}}
$$
\medskip
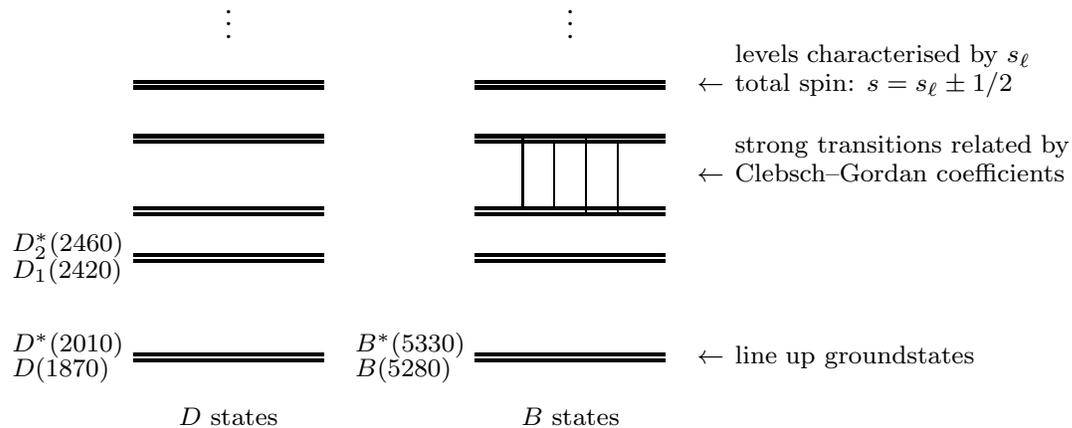
\figcaption{Spectrum of states predicted by heavy quark symmetry}
\bigskip

\newbox\halfglueloop
\setbox\halfglueloop=\hbox{\unit=0.5pt
\cpt 0.0000 46.0000 \cpt 1.1470 45.9442 \cpt 2.2775 45.7778 \cpt
3.3771 45.5032 \cpt 4.4315 45.1242 \cpt 5.4271 44.6456 \cpt
6.3508 44.0735 \cpt 7.1905 43.4151 \cpt 7.9345 42.6787 \cpt
8.5724 41.8736 \cpt 9.0945 41.0100 \cpt 9.4923 40.0994 \cpt
9.7581 39.1546 \cpt 9.8855 38.1901 \cpt 9.8689 37.2231 \cpt
9.7037 36.2744 \cpt 9.3871 35.3718 \cpt 8.9180 34.5540 \cpt
8.3009 33.8784 \cpt 7.5537 33.4316 \cpt 6.7288 33.3308 \cpt
5.9402 33.6709 \cpt 5.3328 34.4066 \cpt 4.9770 35.3672 \cpt
4.8610 36.4081 \cpt 4.9524 37.4502 \cpt 5.2228 38.4522 \cpt
5.6499 39.3903 \cpt 6.2153 40.2491 \cpt 6.9028 41.0176 \cpt
7.6974 41.6873 \cpt 8.5843 42.2518 \cpt 9.5490 42.7060 \cpt
10.5767 43.0462 \cpt 11.6528 43.2699 \cpt 12.7622 43.3758 \cpt
13.8900 43.3642 \cpt 15.0214 43.2363 \cpt 16.1416 42.9947 \cpt
17.2363 42.6434 \cpt 18.2913 42.1875 \cpt 19.2933 41.6332 \cpt
20.2294 40.9880 \cpt 21.0878 40.2604 \cpt 21.8574 39.4600 \cpt
22.5279 38.5969 \cpt 23.0906 37.6823 \cpt 23.5375 36.7280 \cpt
23.8620 35.7464 \cpt 24.0587 34.7506 \cpt 24.1237 33.7541 \cpt
24.0542 32.7712 \cpt 23.8492 31.8176 \cpt 23.5095 30.9102 \cpt
23.0379 30.0690 \cpt 22.4407 29.3183 \cpt 21.7294 28.6905 \cpt
20.9249 28.2308 \cpt 20.0670 28.0047 \cpt 19.2315 28.0980 \cpt
18.5466 28.5815 \cpt 18.1513 29.4122 \cpt 18.0872 30.4164 \cpt
18.2996 31.4380 \cpt 18.7224 32.3968 \cpt 19.3081 33.2573 \cpt
20.0232 34.0024 \cpt 20.8431 34.6234 \cpt 21.7475 35.1152 \cpt
22.7188 35.4754 \cpt 23.7405 35.7029 \cpt 24.7971 35.7984 \cpt
25.8735 35.7635 \cpt 26.9548 35.6012 \cpt 28.0267 35.3156 \cpt
29.0754 34.9123 \cpt 30.0873 34.3976 \cpt 31.0497 33.7793 \cpt
31.9504 33.0662 \cpt 32.7781 32.2680 \cpt 33.5225 31.3953 \cpt
34.1742 30.4598 \cpt 34.7251 29.4736 \cpt 35.1682 28.4496 \cpt
35.4975 27.4009 \cpt 35.7087 26.3412 \cpt 35.7985 25.2843 \cpt
35.7652 24.2443 \cpt 35.6081 23.2351 \cpt 35.3283 22.2709 \cpt
34.9284 21.3659 \cpt 34.4125 20.5345 \cpt 33.7871 19.7920 \cpt
33.0612 19.1549 \cpt 32.2480 18.6422 \cpt 31.3674 18.2771 \cpt
30.4503 18.0903 \cpt 29.5482 18.1228 \cpt 28.7482 18.4254 \cpt
28.1844 19.0339 \cpt 27.9949 19.8991 \cpt 28.2035 20.8584 \cpt
28.7129 21.7601 \cpt 29.4199 22.5343 \cpt 30.2564 23.1584 \cpt
31.1797 23.6267 \cpt 32.1604 23.9396 \cpt 33.1763 24.0995 \cpt
34.2094 24.1105 \cpt 35.2435 23.9778 \cpt 36.2640 23.7074 \cpt
37.2573 23.3064 \cpt 38.2107 22.7827 \cpt 39.1122 22.1455 \cpt
39.9506 21.4047 \cpt 40.7159 20.5712 \cpt 41.3987 19.6568 \cpt
41.9908 18.6738 \cpt 42.4851 17.6356 \cpt 42.8757 16.5557 \cpt
43.1579 15.4482 \cpt 43.3284 14.3275 \cpt 43.3852 13.2081 \cpt
43.3275 12.1044 \cpt 43.1560 11.0310 \cpt 42.8724 10.0018 \cpt
42.4801 9.0307 \cpt 41.9835 8.1309 \cpt 41.3884 7.3153 \cpt
40.7021 6.5962 \cpt 39.9333 5.9855 \cpt 39.0931 5.4949 \cpt
38.1948 5.1359 \cpt 37.2562 4.9208 \cpt 36.3012 4.8628 \cpt
35.3648 4.9776 \cpt 34.5010 5.2829 \cpt 33.7956 5.7956 \cpt
33.3755 6.5134 \cpt 33.3737 7.3698 \cpt 33.8089 8.2154 \cpt
34.5530 8.9173 \cpt 35.4666 9.4289 \cpt 36.4626 9.7491 \cpt
37.4913 9.8882 \cpt 38.5217 9.8578 \cpt 39.5324 9.6690 \cpt
40.5068 9.3323 \cpt 41.4313 8.8588 \cpt 42.2943 8.2592 \cpt
43.0855 7.5453 \cpt 43.7958 6.7290 \cpt 44.4171 5.8233 \cpt
44.9427 4.8412 \cpt 45.3669 3.7965 \cpt 45.6851 2.7037 \cpt
45.8942 1.5772 \cpt 45.9921 0.4318 \cpt 45.9782 -0.7174 \cpt
45.8532 -1.8557 \cpt 45.6189 -2.9685 \cpt 45.2786 -4.0415 \cpt
44.8368 -5.0607 \cpt 44.2991 -6.0128 \cpt 43.6723 -6.8854 \cpt
42.9642 -7.6666 \cpt 42.1838 -8.3456 \cpt 41.3410 -8.9122 \cpt
40.4467 -9.3577 \cpt 39.5132 -9.6740 \cpt 38.5544 -9.8542 \cpt
37.5861 -9.8923 \cpt 36.6278 -9.7835 \cpt 35.7041 -9.5242 \cpt
34.8489 -9.1124 \cpt 34.1114 -8.5500 \cpt 33.5660 -7.8481 \cpt
33.3203 -7.0429 \cpt 33.4897 -6.2232 \cpt 34.0931 -5.5334 \cpt
34.9898 -5.0813 \cpt 36.0126 -4.8786 \cpt 37.0606 -4.8956 \cpt
38.0811 -5.1014 \cpt 39.0453 -5.4717 \cpt 39.9355 -5.9869 \cpt
40.7391 -6.6303 \cpt 41.4469 -7.3862 \cpt 42.0517 -8.2400 \cpt
42.5480 -9.1770 \cpt 42.9315 -10.1826 \cpt 43.1993 -11.2420 \cpt
43.3497 -12.3406 \cpt 43.3824 -13.4632 \cpt 43.2981 -14.5949 \cpt
43.0991 -15.7209 \cpt 42.7887 -16.8269 \cpt 42.3716 -17.8983 \cpt
41.8537 -18.9218 \cpt 41.2419 -19.8842 \cpt 40.5445 -20.7734 \cpt
39.7704 -21.5777 \cpt 38.9297 -22.2868 \cpt 38.0333 -22.8912 \cpt
37.0925 -23.3827 \cpt 36.1199 -23.7541 \cpt 35.1279 -23.9996 \cpt
34.1300 -24.1148 \cpt 33.1403 -24.0964 \cpt 32.1738 -23.9427 \cpt
31.2468 -23.6537 \cpt 30.3780 -23.2315 \cpt 29.5898 -22.6808 \cpt
28.9110 -22.0109 \cpt 28.3813 -21.2383 \cpt 28.0577 -20.3942 \cpt
28.0194 -19.5381 \cpt 28.3516 -18.7783 \cpt 29.0666 -18.2612 \cpt
30.0276 -18.0751 \cpt 31.0556 -18.1912 \cpt 32.0440 -18.5410 \cpt
32.9443 -19.0695 \cpt 33.7345 -19.7388 \cpt 34.4032 -20.5216 \cpt
34.9443 -21.3962 \cpt 35.3546 -22.3441 \cpt 35.6323 -23.3485 \cpt
35.7777 -24.3936 \cpt 35.7919 -25.4642 \cpt 35.6774 -26.5453 \cpt
35.4380 -27.6225 \cpt 35.0787 -28.6815 \cpt 34.6054 -29.7088 \cpt
34.0255 -30.6913 \cpt 33.3473 -31.6167 \cpt 32.5802 -32.4733 \cpt
31.7345 -33.2504 \cpt 30.8216 -33.9383 \cpt 29.8532 -34.5283 \cpt
28.8421 -35.0130 \cpt 27.8013 -35.3861 \cpt 26.7442 -35.6426 \cpt
25.6847 -35.7788 \cpt 24.6367 -35.7924 \cpt 23.6142 -35.6824 \cpt
22.6313 -35.4490 \cpt 21.7021 -35.0944 \cpt 20.8410 -34.6220 \cpt
20.0630 -34.0374 \cpt 19.3838 -33.3484 \cpt 18.8212 -32.5666 \cpt
18.3969 -31.7088 \cpt 18.1380 -30.8010 \cpt 18.0815 -29.8852 \cpt
18.2752 -29.0331 \cpt 18.7665 -28.3612 \cpt 19.5484 -28.0177 \cpt
20.4938 -28.0821 \cpt 21.4296 -28.4908 \cpt 22.2567 -29.1325 \cpt
22.9391 -29.9259 \cpt 23.4669 -30.8201 \cpt 23.8387 -31.7819 \cpt
24.0564 -32.7868 \cpt 24.1236 -33.8153 \cpt 24.0449 -34.8507 \cpt
23.8261 -35.8781 \cpt 23.4739 -36.8833 \cpt 22.9959 -37.8533 \cpt
22.4009 -38.7758 \cpt 21.6983 -39.6396 \cpt 20.8987 -40.4338 \cpt
20.0137 -41.1490 \cpt 19.0554 -41.7765 \cpt 18.0366 -42.3088 \cpt
16.9710 -42.7395 \cpt 15.8724 -43.0634 \cpt 14.7550 -43.2768 \cpt
13.6333 -43.3771 \cpt 12.5220 -43.3630 \cpt 11.4355 -43.2347 \cpt
10.3878 -42.9935 \cpt 9.3930 -42.6422 \cpt 8.4645 -42.1846 \cpt
7.6153 -41.6263 \cpt 6.8581 -40.9738 \cpt 6.2047 -40.2353 \cpt
5.6669 -39.4209 \cpt 5.2564 -38.5429 \cpt 4.9850 -37.6173 \cpt
4.8655 -36.6651 \cpt 4.9124 -35.7166 \cpt 5.1429 -34.8171 \cpt
5.5751 -34.0386 \cpt 6.2178 -33.4926 \cpt 7.0353 -33.3198 \cpt
7.9048 -33.5978 \cpt 8.6713 -34.2430 \cpt 9.2569 -35.1050 \cpt
9.6491 -36.0771 \cpt 9.8558 -37.0977 \cpt 9.8887 -38.1304 \cpt
9.7589 -39.1508 \cpt 9.4773 -40.1407 \cpt 9.0547 -41.0858 \cpt
8.5018 -41.9735 \cpt 7.8300 -42.7932 \cpt 7.0512 -43.5353 \cpt
6.1779 -44.1914 \cpt 5.2233 -44.7542 \cpt 4.2008 -45.2176 \cpt
3.1247 -45.5766 \cpt 2.0093 -45.8276 \cpt 0.8694 -45.9680 }
\def\halfgluonloop{\vcenter{%
\hbox{\copy\halfglueloop}}\hbox to 23.0000pt{\hfil}}
\newbox\smhalfglueloop
\setbox\smhalfglueloop=\hbox{\unit=0.5pt
\cpt 0.0000 26.0000 \cpt 1.2953 25.9262 \cpt 2.5643 25.7069 \cpt
3.7832 25.3476 \cpt 4.9298 24.8571 \cpt 5.9833 24.2469 \cpt
6.9252 23.5308 \cpt 7.7397 22.7247 \cpt 8.4136 21.8458 \cpt
8.9370 20.9125 \cpt 9.3028 19.9433 \cpt 9.5077 18.9569 \cpt
9.5512 17.9719 \cpt 9.4364 17.0062 \cpt 9.1700 16.0779 \cpt
8.7625 15.2059 \cpt 8.2291 14.4124 \cpt 7.5912 13.7271 \cpt
6.8791 13.1950 \cpt 6.1371 12.8877 \cpt 5.4325 12.9065 \cpt
4.8559 13.3342 \cpt 4.4855 14.1222 \cpt 4.3373 15.1053 \cpt
4.3895 16.1451 \cpt 4.6206 17.1652 \cpt 5.0159 18.1240 \cpt
5.5636 18.9955 \cpt 6.2520 19.7608 \cpt 7.0680 20.4049 \cpt
7.9966 20.9155 \cpt 9.0203 21.2824 \cpt 10.1197 21.4978 \cpt
11.2738 21.5565 \cpt 12.4600 21.4559 \cpt 13.6552 21.1967 \cpt
14.8358 20.7824 \cpt 15.9786 20.2200 \cpt 17.0613 19.5195 \cpt
18.0631 18.6941 \cpt 18.9651 17.7594 \cpt 19.7505 16.7337 \cpt
20.4058 15.6371 \cpt 20.9201 14.4913 \cpt 21.2857 13.3187 \cpt
21.4983 12.1424 \cpt 21.5568 10.9849 \cpt 21.4631 9.8681 \cpt
21.2221 8.8126 \cpt 20.8413 7.8372 \cpt 20.3306 6.9583 \cpt
19.7021 6.1903 \cpt 18.9700 5.5445 \cpt 18.1507 5.0298 \cpt
17.2639 4.6529 \cpt 16.3337 4.4187 \cpt 15.3923 4.3327 \cpt
14.4858 4.4032 \cpt 13.6849 4.6460 \cpt 13.0965 5.0853 \cpt
12.8510 5.7326 \cpt 13.0164 6.5293 \cpt 13.5201 7.3482 \cpt
14.2398 8.0873 \cpt 15.0915 8.6962 \cpt 16.0274 9.1508 \cpt
17.0177 9.4387 \cpt 18.0395 9.5533 \cpt 19.0727 9.4921 \cpt
20.0976 9.2558 \cpt 21.0949 8.8486 \cpt 22.0451 8.2775 \cpt
22.9295 7.5528 \cpt 23.7299 6.6876 \cpt 24.4297 5.6978 \cpt
25.0142 4.6018 \cpt 25.4709 3.4203 \cpt 25.7900 2.1756 \cpt
25.9651 0.8915 \cpt 25.9927 -0.4077 \cpt 25.8728 -1.6973 \cpt
25.6089 -2.9531 \cpt 25.2074 -4.1517 \cpt 24.6780 -5.2712 \cpt
24.0330 -6.2916 \cpt 23.2869 -7.1951 \cpt 22.4560 -7.9667 \cpt
21.5580 -8.5944 \cpt 20.6114 -9.0689 \cpt 19.6348 -9.3845 \cpt
18.6470 -9.5386 \cpt 17.6661 -9.5319 \cpt 16.7101 -9.3687 \cpt
15.7973 -9.0567 \cpt 14.9473 -8.6078 \cpt 14.1840 -8.0391 \cpt
13.5409 -7.3741 \cpt 13.0701 -6.6469 \cpt 12.8533 -5.9082 \cpt
12.9954 -5.2333 \cpt 13.5495 -4.7154 \cpt 14.4175 -4.4158 \cpt
15.4299 -4.3332 \cpt 16.4695 -4.4435 \cpt 17.4738 -4.7275 \cpt
18.4075 -5.1717 \cpt 19.2475 -5.7648 \cpt 19.9763 -6.4947 \cpt
20.5798 -7.3477 \cpt 21.0463 -8.3079 \cpt 21.3665 -9.3575 \cpt
21.5332 -10.4763 \cpt 21.5422 -11.6429 \cpt 21.3917 -12.8344 \cpt
21.0834 -14.0274 \cpt 20.6219 -15.1986 \cpt 20.0151 -16.3250 \cpt
19.2739 -17.3845 \cpt 18.4126 -18.3569 \cpt 17.4473 -19.2240 \cpt
16.3973 -19.9701 \cpt 15.2830 -20.5822 \cpt 14.1264 -21.0506 \cpt
12.9503 -21.3687 \cpt 11.7775 -21.5331 \cpt 10.6305 -21.5437 \cpt
9.5308 -21.4033 \cpt 8.4984 -21.1177 \cpt 7.5515 -20.6952 \cpt
6.7059 -20.1463 \cpt 5.9750 -19.4838 \cpt 5.3693 -18.7226 \cpt
4.8971 -17.8801 \cpt 4.5642 -16.9769 \cpt 4.3757 -16.0393 \cpt
4.3374 -15.1029 \cpt 4.4595 -14.2203 \cpt 4.7608 -13.4725 \cpt
5.2658 -12.9775 \cpt 5.9705 -12.8593 \cpt 6.7883 -13.1433 \cpt
7.5904 -13.7263 \cpt 8.2926 -14.4940 \cpt 8.8552 -15.3758 \cpt
9.2588 -16.3314 \cpt 9.4932 -17.3337 \cpt 9.5531 -18.3611 \cpt
9.4371 -19.3937 \cpt 9.1471 -20.4120 \cpt 8.6880 -21.3966 \cpt
8.0679 -22.3283 \cpt 7.2979 -23.1884 \cpt 6.3922 -23.9592 \cpt
5.3673 -24.6245 \cpt 4.2424 -25.1704 \cpt 3.0390 -25.5852 \cpt
1.7795 -25.8601 \cpt 0.4881 -25.9896 }
\def\smhalfgluonloop{\vcenter{%
\hbox{\copy\smhalfglueloop}}\hbox to 13.0000pt{\hfil}}
\def\v{\kern-0.05em$\vee$\kern-0.1em}
\def\WZline{\vcenter{\hbox{\v\v\v\v\v\v\v\v}}}
\def\vprop{\vcenter{\hbox{\vrule height 72pt width1pt}}}
\def\vheavyprop{\vcenter{\hbox{\vrule height72pt width1pt
 \kern0.8pt \vrule height72pt width1pt}}}
\def\uphalfloop{\vcenter{\hbox{$\smhalfgluonloop$}
 \kern32pt\hbox{}}}
\def\dnhalfloop{\vcenter{\hbox{}\kern36pt
 \hbox{$\smhalfgluonloop$}}}
$$
\matrix{\displaystyle
\WZline\vprop + \WZline\vprop\uphalfloop +
 \WZline\vprop\dnhalfloop + \WZline\vprop\halfgluonloop \cr
\noalign{\vskip8pt}
\hbox{Full QCD to order $\alpha_s$} \cr
\noalign{\vskip15pt}
\displaystyle
\WZline\rlap{\kern5.8pt$\hqgamma$}\vheavyprop
 \hbox{\kern3pt$\phantom{\hqgamma}$} +
 \WZline\vheavyprop\uphalfloop +
 \WZline\vheavyprop\dnhalfloop +
 \WZline\vheavyprop\halfgluonloop \cr
\noalign{\vskip8pt}
\hbox{HQET to order $\alpha_s$} \cr}
$$
\setbox\halfglueloop=\hbox{}
\setbox\smhalfglueloop=\hbox{}
\medskip
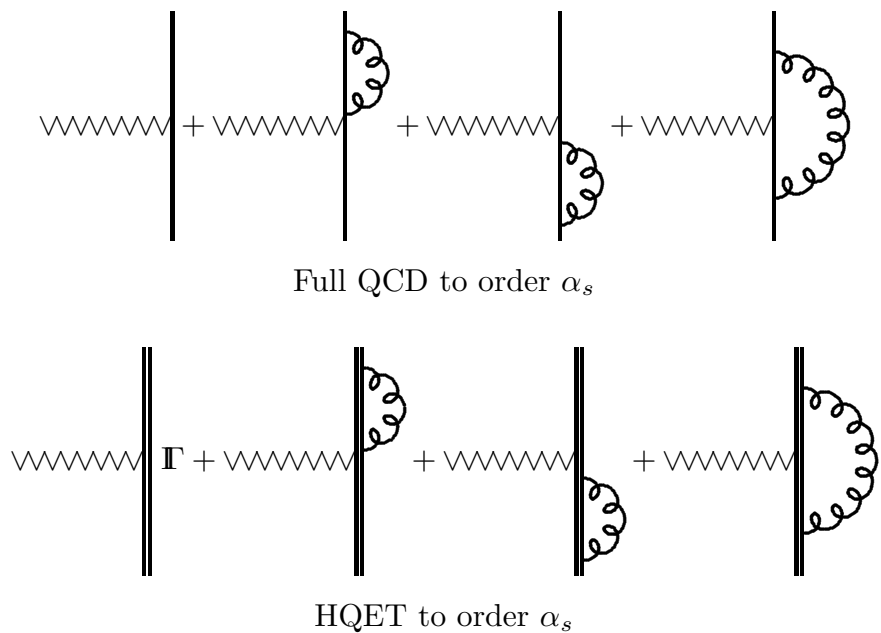
\figcaption{Matching a current between QCD and the heavy quark
effective theory}

\bye